\documentclass[a4paper,11pt]{article}
\pdfoutput=1 
\usepackage{amssymb,mathrsfs}
\usepackage{bbm,bm}
\usepackage{siunitx}
\usepackage{amsmath}
\usepackage{braket}
\usepackage{mathtools} 
\usepackage[usenames,dvipsnames,svgnames,table]{xcolor}
\usepackage [utf8] {inputenc}
\usepackage{color}
\usepackage[pdftex] {graphicx}
\usepackage{subcaption}
\usepackage{adjustbox}
\newcommand{\myfig}[2]{%
	\subcaptionbox{\label{#2}}{%
		\includegraphics[height=3.2cm,keepaspectratio]{#1}%
	}%
	\hspace{0em}%
}
\newcommand{\myfigg}[1]{%
	\begin{minipage}{0.49\linewidth}
		\centering
		\includegraphics[height=5.3cm,keepaspectratio]{#1}
	\end{minipage}%
}
\newcommand{\myfiggg}[1]{%
	\begin{minipage}{0.5\linewidth}
		\centering
		\includegraphics[height=4.9cm,keepaspectratio]{#1}
	\end{minipage}%
}

\usepackage{lmodern}
\usepackage{footnote}
\usepackage{slashed}
\usepackage{mathrsfs}
\usepackage[pdftex] {graphicx}
\usepackage{multirow}
\usepackage{jheppub} 
\usepackage{here}
\usepackage{hyperref}
\usepackage[justification=centering, singlelinecheck=false]{caption}
\usepackage[list=true, labelfont=bf, labelformat=brace, position=top]{subcaption}
\newcommand{\eq}{\begin{eqnarray}}
	\newcommand{\en}{\end{eqnarray}}

\title{Form factors of the $\rho$-meson in chiral perturbation theory}

\renewcommand{\thefootnote}{\fnsymbol{footnote}}

\author[1]{Kakha Shamanauri,}
\affiliation[1]{Bethe Center for Theoretical Physics, Universit\"at Bonn, 53115 Bonn, Germany}
\emailAdd{s53ksham@uni-bonn.de}

\author[2,3]{ Jambul Gegelia,}
\affiliation[2]{Institut f\"ur Theoretische Physik II, Ruhr-Universit\"at Bochum,
	D-44780 Bochum, Germany}
\affiliation[3]{Tbilisi State University, 0186 Tbilisi, Georgia}
\emailAdd{Jambul.Gegelia@ruhr-uni-bochum.de}

\author[4,5]{Ulf-G. Mei{\ss}ner,}
\affiliation[4]{Helmholtz-Institut f\"ur Strahlen- und Kernphysik (Theorie)\\ and Bethe Center for Theoretical Physics, Universit\"at Bonn, 53115 Bonn, Germany}
\affiliation[5]{Institute for Advanced Simulation (IAS-4) Forschungszentrum J{\"u}lich, Germany}
\emailAdd{meissner@hiskp.uni-bonn.de}

\author[4]{Akaki Rusetsky}
\emailAdd{rusetsky@hiskp.uni-bonn.de}

\abstract{
	We present the calculation of the electromagnetic form factors of the $\rho$-meson
	at one loop in Chiral Perturbation Theory. The power-counting-violating terms in the loop diagrams are
	subtracted by using infrared regularization, and the Ward identities are
	explicitly verified at the order considered. The results are compared to the recent
	calculations carried out in the framework of  non-relativistic effective field
	theory.
	
	\noindent
	
}

\allowdisplaybreaks
\begin{document}
	\maketitle
	\flushbottom
	
	\renewcommand{\thefootnote}{\arabic{footnote}}

	\section{Introduction}
	
	The electromagnetic form factors represent
	a useful source of information about the internal
	structure of a hadron. This statement applies equally to the stable hadrons as well as
	to resonances, albeit, for obvious reasons, calculations in case of the
	latter represent a bigger challenge.
	It is particularly interesting that the resonance form factors, unlike those of the
	stable particles, demonstrate peculiar behavior for small momentum transfer.
	In the case of the $\rho$-meson, considered here, that manifests
	itself for instance in the large value of the quadrupole momentum~\cite{Meissner:2026zos}. 
	In our opinion, firmly establishing the slopes of the
	form factors near $q^2=0$ would provide a nice test on the theoretical methods and
	assumptions that are used to describe unstable hadrons at present.
	
	In quantum field theory, resonance form factors are rigorously defined through the
	residue of the five-point Green function at the double pole in the complex energy 
	plane (see, e.g.,~\cite{Gegelia:2010nmt,Djukanovic:2013mka,Bernard:2012bi,RuizdeElvira:2017aet}).\footnote{For simplicity, here it is assumed that
		resonances carry the
		quantum numbers of a two-particle scattering state. The $\rho$-meson, which emerges
		in the $\pi\pi$ scattering in the channel with quantum numbers $J^{PC}=1^{++}$
		is, of course, a prominent example of such a resonance.}
	This is a straightforward generalization of the well-known Mandelstam
	approach~\cite{Mandelstam:1955sd} to the definition of the form factors of 
	stable bound states. On the lattice, the complications arise because the infinite-volume
	limit in this five-point function cannot be performed in a standard manner.
	The problem resides in the so-called triangle diagram (the diagram in which the photon
	is attached to one of the constituents, whereas the other plays a role of a spectator), which
	does not possess a well-defined infinite-volume limit. Two
	solutions have been suggested in the past. In particular, in
	Refs.~\cite{Hoja:2010fm,Bernard:2012bi,Baroni:2018iau,Briceno:2019nns,Briceno:2020xxs,Moscoso:2026wmz}
	it is proposed to subtract the triangle diagram from the finite-volume matrix element,
	calculated on the lattice, and perform the infinite-volume limit in the remaining
	short-distance part of this matrix element.
	An alternative proposal~\cite{Lozano:2022kfz,Meissner:2026zos}
	is based on placing the two-particle system in the spatially periodic external field.
	The calculated energy spectrum can be then directly fitted to the short-distance part
	at different values of the momentum transfer. It remains to be seen, which
	of these two methods will prove to be more convenient for the extraction of the form
	factors of unstable states on the lattice. 
	
	At present, we are not aware of any lattice
	calculation of the form factors of unstable particles
	in the literature. The single existing calculation of the form factors of the $\rho$-meson
	is carried out for the case of a stable $\rho$~\cite{QCDSF:2008tjq}
        (for a review on hadron resonances in lattice QCD, see~\cite{Mai:2022eur}).
		In order to observe
	the predicted large curvature of the form factors near $q^2=0$, however, it is essential to
	carry out calculations at smaller values of the quark masses, when the $\rho$-meson
	moves close to the threshold from below and eventually becomes unstable. In anticipation of such a calculation, which might
	be very challenging, any prior information about the expected behavior of the form
	factors would be very helpful. As already mentioned, in Ref.~\cite{Meissner:2026zos}
	the form factors were evaluated in non-relativistic effective field theory (NREFT),
	and rather robust predictions about the behavior
	of these form factors near $q^2=0$ were made. In this paper, we calculate the same
	form factors in the covariant chiral perturbation theory (ChPT), with the aim to gain
	more information from matching
	our results to those of the NREFT calculations. This is a primary motivation  for
	the present work.
	
\begin{sloppypar}
        An additional important aspect of the calculations in the relativistic framework also has
	to be mentioned. As is well known, loop corrections in the theory with heavy massive particles
	(like the $\rho$-meson) break chiral counting rules and special prescriptions on top
	of the standard dimensional regularization should be applied to eliminate the
	power-counting-breaking polynomial contributions. For example, in
	Ref.~\cite{Djukanovic:2013mka}, the so-called complex-mass renormalization
scheme~\cite{Bauer:2012at,Stuart_in_Z0,Denner:1999gp,Denner:2006ic,Denner:2005fg,Actis:2006rc,Actis:2008uh,Djukanovic:2009zn,Djukanovic:2015gna}
	has been used. In current work we follow mainly
	the methodology outlined in Refs.~\cite{Bruns:2004tj,Bruns:2008ub}, where the
	well-known infrared regularization scheme~\cite{Becher:1999he} is adapted for the theory
	containing massive vector mesons and baryons. We check explicitly that
	the used prescription does not violate Ward identities at the order we are working,
	and, hence, $U(1)$ gauge symmetry stays intact.
        \end{sloppypar}
	
	The layout of the paper is as follows. In Sect.~\ref{sec:dimreg} we display the
	chiral Lagrangian used in calculations, discuss the counting rules and give the result
	of the calculation of the full set of one-loop diagrams in standard dimensional regularization.
	In Sect.~\ref{sec:subtract} we discuss in detail the infrared regularization and display
	subtraction polynomials for each diagram. Sect.~\ref{sec:Ward} contains the discussion
	of the $U(1)$ Ward identity. In Sect.~\ref{sec:numerics} we display the results of the
	numerical calculations and a comparison to Ref.~\cite{Meissner:2026zos}.
	Section~\ref{sec:concl} contains our conclusions. In order to render the paper both
	self-contained and readable, explicit expressions for individual diagrams are relegated
	to the appendices.
	
	\section{The Lagrangian, counting rules and the amplitude}
	\label{sec:dimreg}

	\subsection{The Lagrangian}
	\label{sec:Lagrangian}
	
	There are many equivalent ways to include
	heavy vectors in the mesonic ChPT. We will be using the vector meson dominance
	model, or the model III Lagrangian from Ref.~\cite{Ecker:1989yg}
	(see also Refs.~\cite{Djukanovic:2013mka,Djukanovic:2014rua}). A more general discussion
	of vector mesons in chiral Lagrangian can be found in Ref.~\cite{Meissner:1987ge}.
	In the parameterization used here,
	the vector fields of the $\rho$-mesons transform inhomogeneously under chiral transformations,
	just as in the parameterization adopted by Weinberg in Ref.~\cite{Weinberg:1968de}. Moreover,
	we will be including the $\omega$-meson since the contribution of the
	$\rho^0\omega\pi$ interaction is not negligible, see e.g. Ref.~\cite{Gell-Mann:1962hpq}. 
	
	At the chiral order we are working, the most general Lagrangian describing the pions $(\pi^a)$, the $\rho$-mesons $(\rho^a)$ and the $\omega$-meson is composed of the following pieces:
	\eq\label{eq:L-full}
	\mathcal{L}=\mathcal{L}_\pi+\mathcal{L}_{\rho \pi}+\mathcal{L}_\omega+\mathcal{L}_{\omega \rho \pi}+\cdots\, .
	\en
	The individual expressions read
	\eq
	\mathcal{L}_\pi &=&\frac{F^2}{4} \operatorname{Tr}\left[\partial_\mu U\left(\partial^\mu U\right)^{\dagger}\right]+\frac{F^2 M^2}{4} \operatorname{Tr}\left(U^{\dagger}+U\right)\,, \nonumber \\[2mm]
	\mathcal{L}_{\rho \pi}&=&-\frac{1}{2} \operatorname{Tr}\left(\rho_{\mu \nu} \rho^{\mu \nu}\right)+\left[m_{\rho,0}^2+\frac{c_{x,0} M^2 \operatorname{Tr}\left(U^{\dagger}+U\right)}{4}\right] 
	\operatorname{Tr}\left[\left(\rho^\mu-\frac{i \Gamma^\mu}{g_0}\right)\left(\rho_\mu-\frac{i \Gamma_\mu}{g_0}\right)\right]\,, \nonumber\\[2mm]
	&+& i\,d_x\,\mathrm{Tr}\left(\rho^{\mu\nu}\Gamma_{\mu\nu}\right)
	-\frac{f_V}{\sqrt{2}}\,\mathrm{Tr}\left(\rho_{\mu\nu}f_+^{\mu\nu}\right)
	+i\,\frac{h_V}{m_{\rho,0}^2} \mathrm{Tr}\left(f_+^{\mu\nu}\left[\rho_{\mu\alpha},\rho_{\nu}^{\,\,\,\alpha}\right]\right)\,,\nonumber\\[2mm]
	\mathcal{L}_\omega &=&-\frac{1}{4}\left(\partial_\mu \omega_\nu-\partial_\nu \omega_\mu\right)\left(\partial^\mu \omega^\nu-\partial^\nu \omega^\mu\right)+\frac{m_\omega^2\, \omega_\mu \omega^\mu}{2}\,, \nonumber\\[2mm]
	\mathcal{L}_{\omega \rho \pi} &=& \frac{1}{2} \,g_{\omega \rho \pi}\,\epsilon_{\mu \nu \alpha \beta} \,\omega^\nu \operatorname{Tr}\left(\rho^{\alpha \beta} u^\mu\right)\,,
	\en
	where 
	\eq
	U &=& u^{2}=\exp\!\left(\frac{i\,\boldsymbol{\tau}\!\cdot\!\boldsymbol{\pi}}{F}\right)\,,\nonumber\\[2mm]
	\rho^{\mu} &=& \frac{\boldsymbol{\tau}\!\cdot\!\boldsymbol{\rho}^{\,\mu}}{2}\,,\nonumber\\[2mm]
	\rho^{\mu\nu} &=& \partial^{\mu}\rho^{\nu}-\partial^{\nu}\rho^{\mu}-i g\,\big[\rho^{\mu},\rho^{\nu}\big]\,,\nonumber\\[2mm]
	u_{\mu} &=& i\!\left[u^{\dagger}\partial_{\mu}u-u\,\partial_{\mu}u^{\dagger}
	-i\!\left(u^{\dagger}v_{\mu}u-u\,v_{\mu}u^{\dagger}\right)\right]\,,\nonumber\\[2mm]
	\Gamma_{\mu} &=& \frac{1}{2}\!\left[u^{\dagger}\partial_{\mu}u+u\,\partial_{\mu}u^{\dagger}
	-i\!\left(u^{\dagger}v_{\mu}u+u\,v_{\mu}u^{\dagger}\right)\right]\,,\nonumber\\[2mm]
	\Gamma_{\mu\nu} &=& \partial_{\mu}\Gamma_{\nu}-\partial_{\nu}\Gamma_{\mu}
	+\big[\Gamma_{\mu},\Gamma_{\nu}\big]\,,\nonumber\\[2mm]
	f_{+}^{\mu\nu} &=& u F^{\mu\nu}u^{\dagger}+u^{\dagger}F^{\mu\nu}u\,,\nonumber\\[2mm]
	F_{\mu\nu}  &=& \partial_\mu v_\nu-\partial_\nu v_\mu  \,,\nonumber\\[2mm]
	D_{\mu}X &=& \partial_{\mu}X-i v_{\mu}X+i X v_{\mu}\, ,
	\en
	and the (external) electromagnetic field enters through $v_\mu=-e \tau^3A_\mu/2$, where
	$e$ is the elementary charge.
	Furthermore, $F$ denotes the pion decay constant in the chiral limit, $M$ is the
	lowest order pion mass and the (real) parameters $m_{\rho,0}$, $m_{\omega}$ are
	the lowest-order $\rho$- and $\omega$-meson masses, respectively.
	The $\rho$-meson couples with the same strength to the pions as with
	itself~\cite{Sakurai:1960ju,Meissner:1987ge,Djukanovic:2004mm} and hence we
	have one coupling $g_0$ for both types
	of interactions. We use the subscript 0 for the {\em bare} quantities $m_{\rho,0}$, $c_{x,0}$,
	and $g_0$ to indicate that only these parameters are renormalized, while all other parameters
	are taken to have their physical values in the chiral limit.\footnote{In what follows, we shall also be using the quantities $m_{\rho}, c_{x},g$, i.e. without the subscript 0, which are the corresponding physical values in the chiral limit.} Moreover, the consistency
	of the theory
	requires that we impose the KSFR relation~\cite{Kawarabayashi:1966kd,Riazuddin:1966swq} on the
	{\em bare} couplings~\cite{Djukanovic:2004mm}
	\eq
	m_{\rho,0}^2=2\,g_0^2F^2.
	\en
	The coupling to the $\omega$-meson is denoted by $g_{\omega \rho \pi}$. Finally, all the
	fields appearing in the Lagrangian are {\em bare} fields. More general discussions of
	the realization of the KSFR relation and the role the $\omega\rho\pi$ coupling
	can be found in Ref.~\cite{Meissner:1987ge}.
	
	Further, we include higher order operators with couplings
	$c_x$, $d_x$, $f_V$ and $h_V$. Note that these {\em should be}
	there {\it a priori}, since they are needed both for the removal of the UV divergences as
	well as the power-counting-violating polynomial terms. Furthermore,
	some of them feature direct $\rho\rho\gamma$ coupling. If the vector meson dominance
	is a good approximation for the $\rho$-meson form factors,  {\em finite parts}  of these
	effective couplings should be small~\cite{Djukanovic:2014rua}. We shall address this
	issue explicitly in the following.

	\subsection{Counting rules}
	
	Let $q$ denote a generic small parameter in the chiral expansion (e.g.,
	the pion mass $M=O(q)$).\footnote{Note that $q$ also denotes the incoming
		photon momentum in the following, which is also assumed to be small.}
	Following closely  
	Ref.~\cite{Djukanovic:2009zn} (see also~\cite{Bruns:2004tj}), the {\em (naive)}
	power counting rules
	are defined as follows. First, the $\rho$- and $\omega$-meson masses are treated as a
	hard scale of the theory and therefore assigned the chiral order $O(1)$.
	At the Lagrangian level, the derivatives acting on the heavy vector mesons
	are considered  as $O(1)$, while the derivatives acting on the pion and the
	electromagnetic fields are treated as $O(q)$. 
	For the momentum
	$p$ in the vicinity of the vector meson mass shell, we take $p^2-m_V^2=O(q)$,
	where $m_V,~V=\rho,\omega$ denotes the heavy vector mass parameter in the Lagrangian.
	In addition, we set
	$m_\omega^2-m_\rho^2\doteq\Delta=O(q^2)$.  
	The electric charge $e$ is counted  as  order $q$, see e.g. Ref.~\cite{Epelbaum:2004fk}.
	
	Furthermore, we distinguish   between the large $p=O(1)$ and small $q=O(q)$
	momenta flowing through the  propagators
	\eq
	D^{\mu\nu}_V(p)=\frac{-g^{\mu\nu}+\dfrac{p^\mu p^\nu}{m_V^2}}{m_V^2-p^2}&=&O(q^{-1}),\qquad D^{\mu\nu}_V(q)=\frac{-g^{\mu\nu}+\dfrac{q^\mu q^\nu}{m_V^2}}{m_V^2-q^2 }=O(1),\nonumber\\[2mm]
	D_{\pi}(p)=\frac{1}{M^2-p^2}&=&O(1),\qquad
	D_{\pi}(q)=\frac{1}{M^2-q^2}=O(q^{-2})\, .
	\en
	The counting of the various vertices emerging from the Lagrangian is straightforward.
	In order to assign a chiral order to a given diagram, one has to consider all
	different ways the large external momenta can flow through it. The lowest order
	that we find in this manner will then give the chiral order of the diagram.

	\begin{figure}[t]
		\centering
		\includegraphics[width=0.4\linewidth]{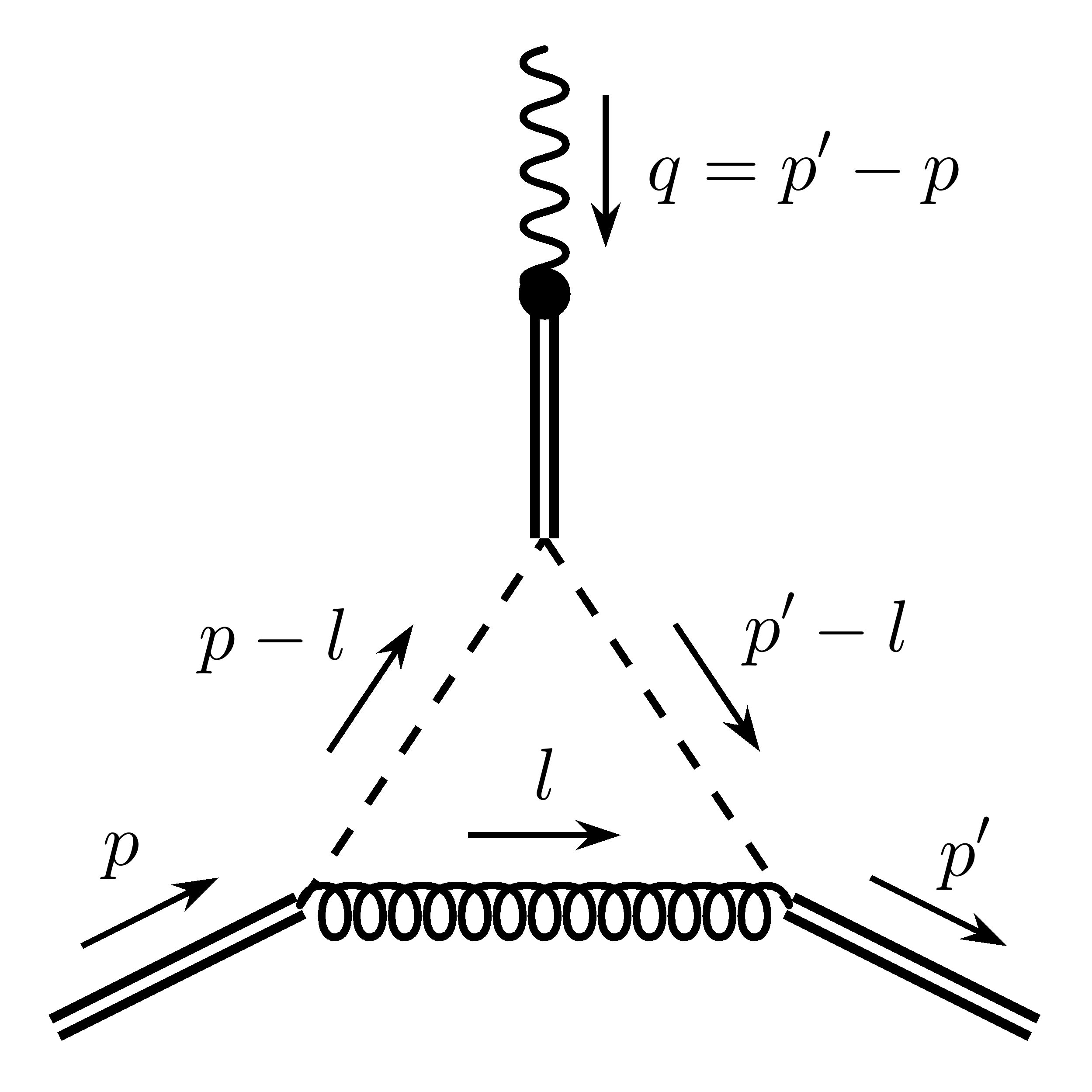}
		\caption{One of the diagrams contributing to the form factors. Wavy, squiggly,
			dashed and  double lines correspond to the photon, $\omega$, $\pi$ and $\rho$-mesons,
			respectively. External $\rho$-meson lines are on-shell, while the photon is off-shell.}
		\label{fig:PowerCounting}
	\end{figure}
	
	Let us illustrate the procedure for the triangle graph,
	shown in Fig.~\ref{fig:PowerCounting}. 
	There exist two different routings of large momenta. First, the large momentum
	from the incoming on-shell $\rho$-meson can flow through the pion line, that is,
	$l=O(q)$, $p-l=O(1)$ and $p'-l=O(1)$.  The  $\gamma\rho$ vertex is counted as
	$O(q)$, because it carries the electric charge. Both $\rho \omega\pi$ vertices are
	$O(1)$, since both external $\rho$ as well as the internal pion lines are heavy (the latter
	because of the large momentum flowing trough). For
	the same reason,  $\rho \pi\pi$ vertex also counts as $O(1)$. All propagators
	in this case contribute at order $O(1)$. The loop integration gives $O(q^4)$.
	Finally, for the given configuration, we obtain that the diagram counts as $O(q^5)$.
	
	For another routing of momenta, namely, when the large external momentum coming
	from the $\rho$-meson flows through the $\omega$ line, the pion propagators
	contribute at order $q^{-2}$ each. Furthermore, the $\omega$-propagator counts  as 
	order $q^{-1}$. Since the pion lines are soft now, all three vertices count  as $O(q)$.
	In this case, one finds that the whole diagram counts as $O(q^3)$. Finally, choosing
	the lowest power for two different routings, we assign the chiral order $O(q^3)$ to
	this diagram.
	
	In reality, it is well known that the naive power counting is violated because of
	the presence of the heavy scale in the propagators. For example, the particular
	diagram in Fig.~\ref{fig:PowerCounting} actually contributes at order $O(q)$ and not
	at $O(q^3)$. This can be straightforwardly checked, performing calculations
	in dimensional regularization. Schematically, the  result can be written as
	(for the full expression see appendix~\ref{app:individual}):
	\eq
	I=e(a_3I_3+a_7I_7+a_8I_8+a_{11}I_{11}+a_{13}I_{13})\, ,
	\en
	where $I_\alpha$ with $\alpha=1,\ldots,13$ are the scalar integrals defined below in
	Eq.~(\ref{eq:Ia}),
	and the coefficients $a_\alpha$ are rational functions of momenta, masses and the spacetime
	dimension $D$. Out of these, only $a_{13}$ is of order $O(1)$, the rest counts
        at higher
	chiral order. The integral $I_{13}$ is also $O(1)$ since it corresponds to a tadpole with
	the mass $m_\omega$. Since $e=O(q)$, we finally conclude that the whole diagram
	contributes at $O(q)$.
	Of course, it is known that the terms that violate power counting are polynomials  of momenta
	and $M^2$ and
	can be removed by adjusting the renormalization prescription.
	Below, in Sect.~\ref{sec:subtract} it will be demonstrated, how this goal can be achieved
	with the use of the so-called infrared regularization (IR)~\cite{Becher:1999he}.
	
	\subsection{Resonance matrix element and kinematics}
	
	Below, we shall rigorously define the notion of the resonance matrix element
	(we remind the reader that a resonance does not correspond to a state vector in the
	Fock space and therefore the matrix element cannot be defined in the usual manner).
	The definition given below represents a generalization of the Mandelstam's approach
	to the stable bound states~\cite{Mandelstam:1955sd} and follows
	closely Refs.~\cite{Meissner:2026zos,Bernard:2012bi,Gegelia:2010nmt,Djukanovic:2013mka,Djukanovic:2009zn}. We start with the two-point function of the $\rho$-meson field
	which is given by\footnote{For simplicity we are suppressing the isospin index
		$a=1,\,2,\,3,$ with respect to which the two-point function is diagonal.}
	\eq
	D_{\mu\nu}(p)=i\int d^4 x \,e^{i p x}\, \langle0|T\rho_{\mu}(x)\rho^{\dagger}_{\nu}(0)|0\rangle\,.
	\en
	The resonance is located at  $p^2=s_R$ that corresponds to a simple pole on
	the second Riemann sheet in the complex variable $p^2$. Near the resonance
	pole, the two-point function behaves as 
	\eq\label{eq:Two-Point Function near resonance}
	D_{\mu\nu}(p)\rightarrow \frac{Z_{\rho}}{s_R-p^2}\sum_{s} \epsilon_{\mu}(p,s)
	\Tilde{\epsilon}_{\nu}(p,s)=\frac{Z_{\rho}}{s_R-p^2}\left(-g_{\mu\nu}+\frac{p_{\mu}p_{\nu}}{s_R}\right),
	\en
	where $Z_{\rho}$ is the  residue of the propagator, $\epsilon_{\mu}(p,\lambda)$ denotes the
	polarization vector and $\Tilde{\epsilon}_{\mu}(p,\lambda)$ is its conjugate. We shall
	follow the definitions of Ref.~\cite{Meissner:2026zos}
	\eq
	\epsilon_{\mu}(p,\pm 1)=\frac{1}{\sqrt{2}}\begin{pmatrix}
		0\\
		\mp 1 \\
		-i \\
		0
	\end{pmatrix}, \qquad  \epsilon_{\mu}(p,0)=\frac{1}{\sqrt{s_R}}\begin{pmatrix}
		p_3\\
		0 \\
		0 \\
		\sqrt{p^2+s_R}
	\end{pmatrix},
	\en
	and define the conjugates via 
	\eq
	\Tilde{\epsilon}_{\mu}(p,\pm 1)=\epsilon^*_{\mu}(p,\pm 1), \qquad
	\Tilde{\epsilon}_{\mu}(p,0)=\epsilon_{\mu}(p,0).
	\en
	It is easy to see that these satisfy the usual properties of the polarization vectors, namely, they
	are transverse $p^{\mu}\epsilon_{\mu}(p,\lambda)=p^{\mu}\Tilde\epsilon_{\mu}(p,\lambda)=0$, orthonormal $\epsilon^{\mu}(p,\lambda)
	\Tilde{\epsilon}_{\mu}(p,\lambda')=-\delta_{\lambda\lambda'}$ and complete in the sense of
	Eq.~(\ref{eq:Two-Point Function near resonance}).
	
	The self-energy of the $\rho$-meson takes the form
	\eq\label{eq:SE}
	\, \Pi_{\mu\nu}(p)
	= 
	g_{\mu\nu}\Pi_1(p^2)
	+ p_\mu p_\nu \Pi_2(p^2),
	\en
	and the residue of the propagator is given by
	\eq\label{eq: Zfactor}
	Z_{\rho}=\frac{1}{1-\Pi'_1(s_R)}.
	\en
	Next, we consider the three-point function
	\eq
	G_{\alpha\beta}^{\lambda}(p',p)
	=
	\int d^4x\, d^4y\,
	e^{ip'x-ipy}
	\langle 0 | T \rho_{\alpha}(x) J^\lambda(0) \rho^{\dagger}_{\beta}(y) | 0 \rangle .
	\en
	Here, $J^{\lambda}(0)$ is the electromagnetic current. Furthermore, it is possible
	to single out the double-pole contribution in this  three-point function, as ${p'}^2,p^2\to s_R$
	\eq\label{eq:Vertex definition}
	G_{\alpha\beta}^{\lambda}(p',p)
	\rightarrow
	-\sqrt{Z_{\rho}}\,
	\frac{-g_{\alpha\mu}+p'_{\mu}p'_{\alpha}/s_R}{s_R-{p'}^2}\,
	\Gamma^{\lambda\mu\nu}(p',p)\,
	\frac{-g_{\nu\beta}+p_{\nu}p_{\beta}/s_R}{s_R-p^2}\,
	\sqrt{Z_{\rho}}.
	\en
	Here, $\Gamma^{\lambda\mu\nu}(p',p)$ is the renormalized
	$\gamma\rho\rho$ vertex,
	which is put on mass shell ${p'}^2=p^2=s_R$. It is useful to think about the above
	formula as a subgraph contributing to the resonant part of the pion-pion scattering
	$\pi\pi\rightarrow\pi\pi\gamma$. We further define the resonance matrix element
	by\footnote{The l.h.s. in this formula is merely a notation since, as already mentioned,
		there are no corresponding one-particle states in the Fock space.}
	\eq
	\langle p',s' | J^\lambda(0) | p,s \rangle\, \doteq\, 
	\Tilde{\epsilon}_{\nu}(p',s')\Gamma^{\lambda\mu\nu}(p',p)\epsilon_{\mu}(p,s)\,.
	\en
	We can also express the above matrix element in terms of the amputated Green function
	\eq
	\label{eq:Matrix element}
	\langle p',s' | J^\lambda(0) | p,s \rangle = -\lim_{{p'}^2, \,\, p^2\rightarrow s_R}
	\Tilde{\epsilon}^{\nu}(p',s')\sqrt{Z_{\rho}}\,G_{\mu\nu;\,\textrm{amp}}^{\lambda}(p',p)\sqrt{Z_{\rho}}\,\epsilon^{\mu}(p,s)\, ,
	\en
	where
	\eq\label{eq:Amputate}
	G_{\mu\nu;\,\textrm{amp}}^{\lambda}(p',p)=(D^{-1})_\mu^\alpha(p)
	G_{\alpha\beta}^{\lambda}(p',p)(D^{-1})_\nu^\beta(p')\, .
	\en
	The vertex $\Gamma^{\lambda\mu\nu}(p',p)$ can be decomposed into three
	independent tensor structures
	\eq\label{eq:Form Factors}
	\Gamma^{\lambda\mu\nu}(p',p)
	=
	-f_1(q^2) g^{\mu\nu}(p+p')^\lambda
	&-&f_2(q^2)\left(q^\nu g^{\lambda\mu}-q^\mu g^{\lambda\nu}\right)\nonumber\\[2mm]
	&+&\frac{1}{2s_R}f_3(q^2)q^\mu q^\nu(p+p')^\lambda ,
	\en
	where the (complex-valued) quantities $f_{1,2,3}(q^2)$ represent the $\rho$-meson form
	factors.

	\subsection{Form Factors}
	In this section, we discuss the electromagnetic form factors up to one-loop  order.
	From the previous discussion, namely Eqs.~(\ref{eq:Matrix element}) and (\ref{eq:Amputate}),
	we can expand the form factors in a perturbative series
	\eq\label{eq:tree-loop}
	f_i(q^2)=f_i^{\textrm{tree}}(q^2)(1+\delta Z_\rho) +f_i^{\textrm{loop}}(q^2)+
	\cdots\, ,
	\en
	where $f_i^{\textrm{tree}}(q^2)$ and $f_i^{\textrm{loop}}(q^2)$  correspond to the tree and
	one-loop order expressions of the three-point vertex function, respectively,
	and $\delta Z_\rho$ is the one-loop order contribution to the residue of the
	propagator, $Z_\rho=1+\delta Z_\rho + \cdots$. 
	
	A brief comment about the power counting is in order.
	We are calculating the amplitude $\Gamma^{\lambda\mu\nu}$ appearing in
	Eq.~(\ref{eq:Form Factors}) up-to-and-including $O(q^3)$. Then, it is clear that the  
	form factors $f_{1}(q^2),f_{2}(q^2),f_{3}(q^2)$ are reliably calculated up to
	and including $O(q^3),O(q^2),O(q)$, respectively (we remind the reader that
	the electric charge $e$ counts as $O(q)$).\footnote{Our full Lorentz-invariant expressions contain strings of higher-order terms as well, since a systematic expansion in powers of $q$ is not always convenient.}
	Furthermore, since the tree  diagrams start at order $O(q)$
		it is clear from Eq.~(\ref{eq:tree-loop}) that
	$\delta Z_\rho$ up-to-and-including $O(q^2)$ is needed. Since, in turn,
	$\delta Z_\rho$ is obtained from the expansion of the  self-energy near the mass shell, 
	\eq
	\Pi(p^2) = \Pi(s_R) +(p^2 - s_R) \Pi'(s_R) + O\!\left((p^2 - s_R)^2\right) ,
	\en
	one may conclude that the quantity $\Pi(p^2)$ is needed up-to-and-including
	$O(q^3)$.        
	
\begin{figure}[htbp]
		\centering
		\myfig{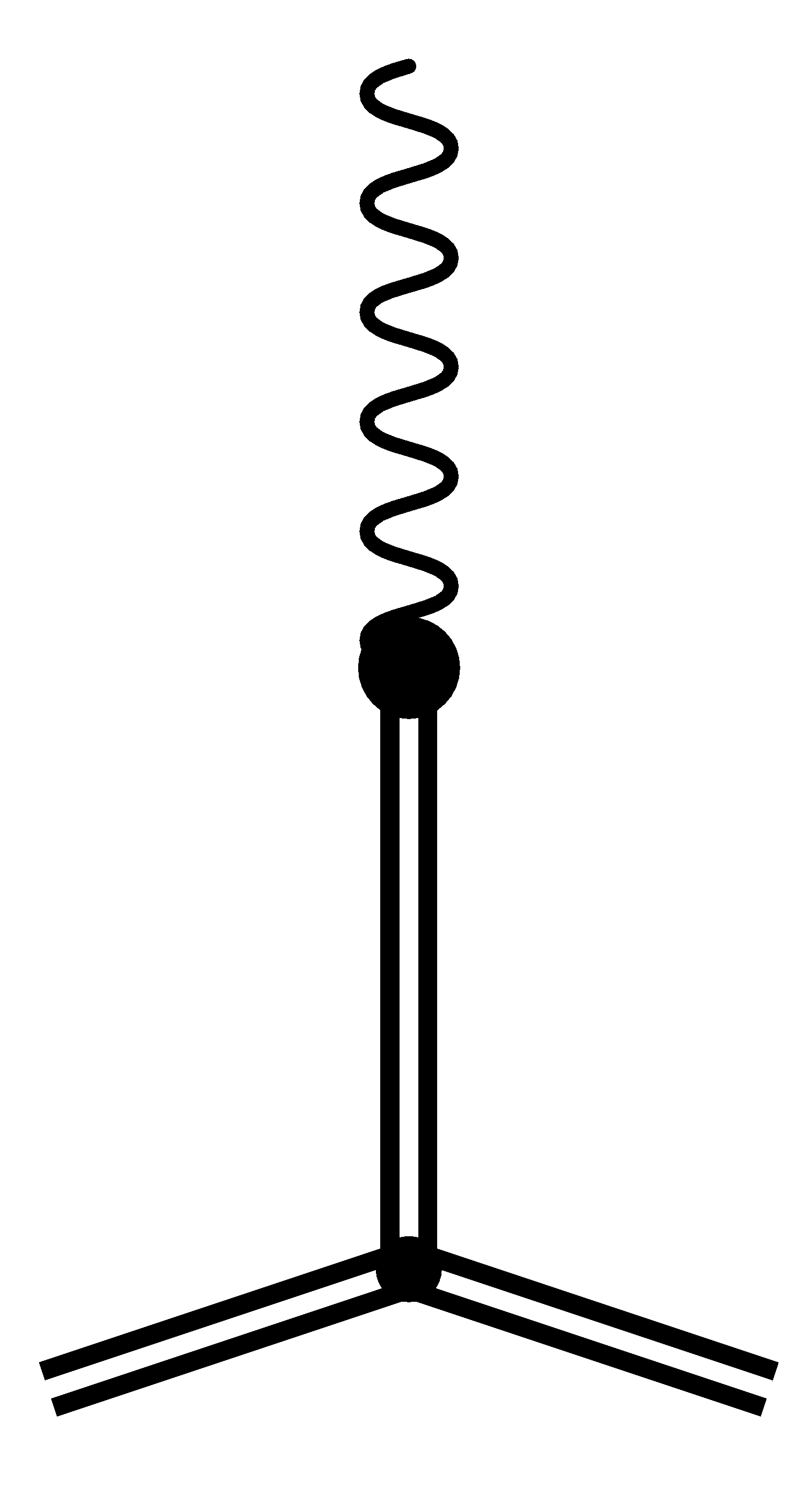}{fig:Tree1}
		\hspace{3cm}
		\myfig{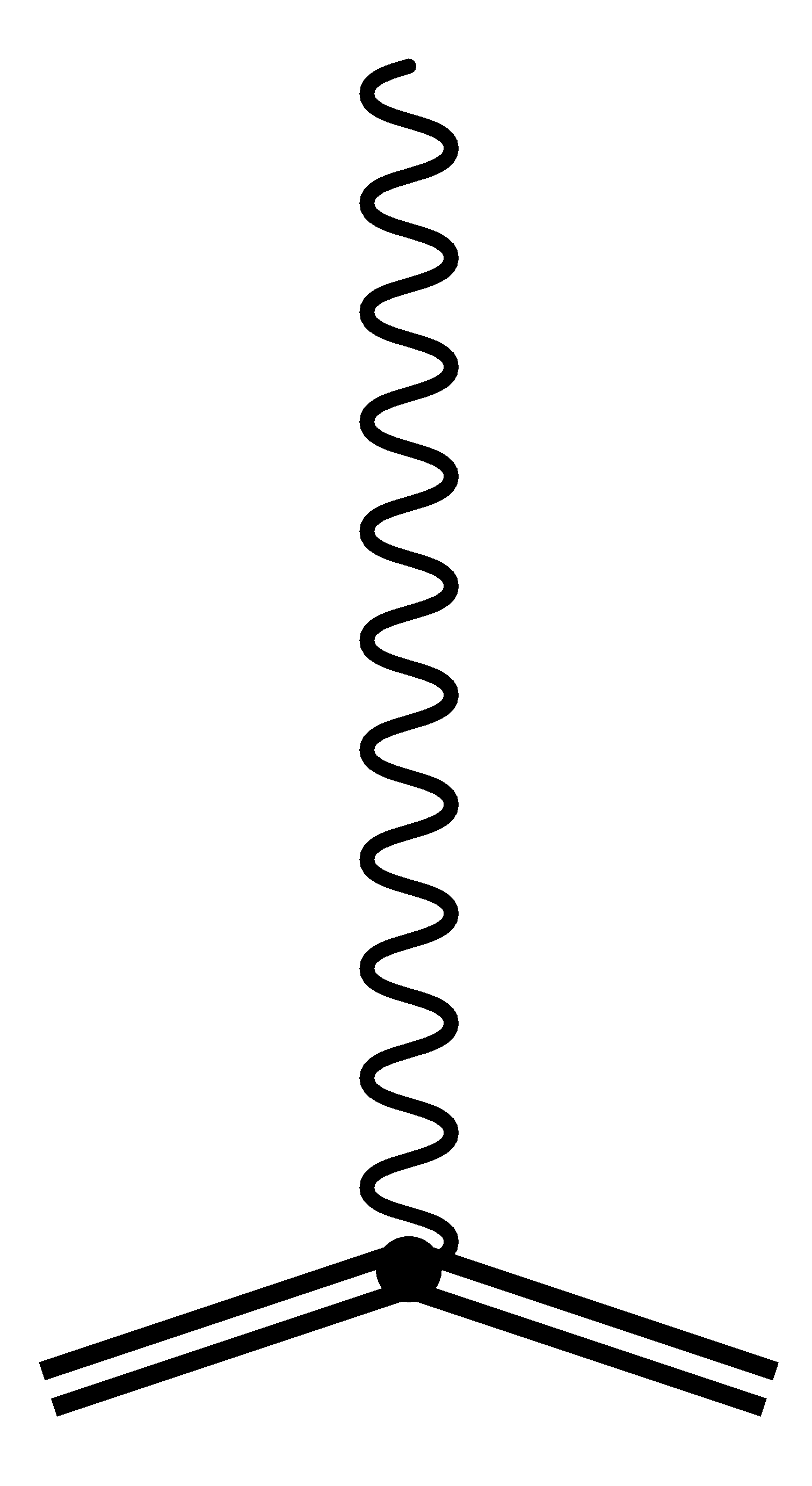}{fig:Tree2}
		
		\myfig{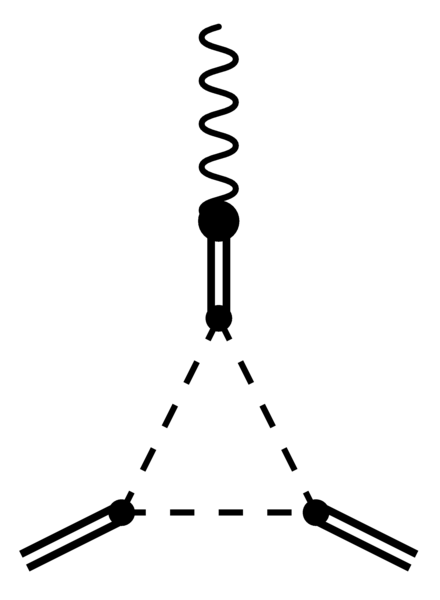}{fig:Loop1}
		\myfig{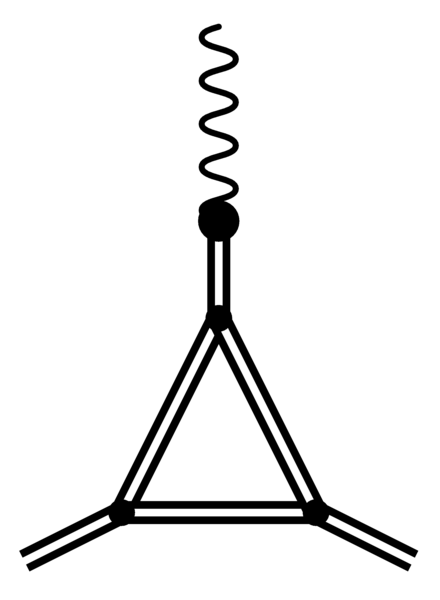}{fig:Loop2}
		\myfig{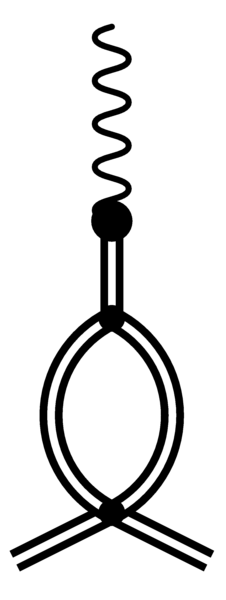}{fig:Loop4}
		\myfig{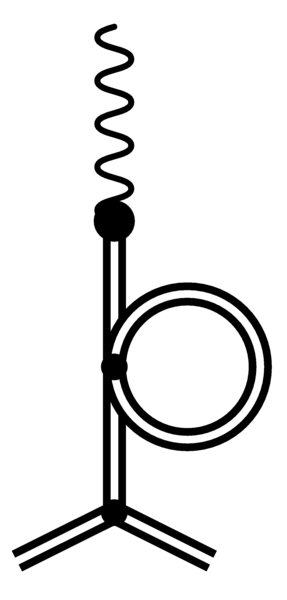}{fig:Loop6}
		\myfig{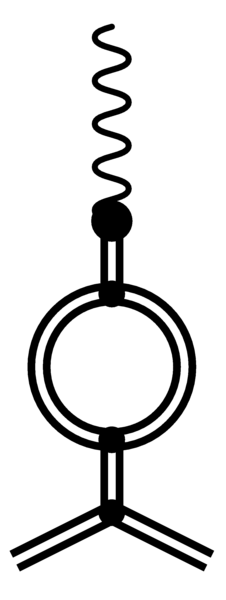}{fig:Loop7}
		\myfig{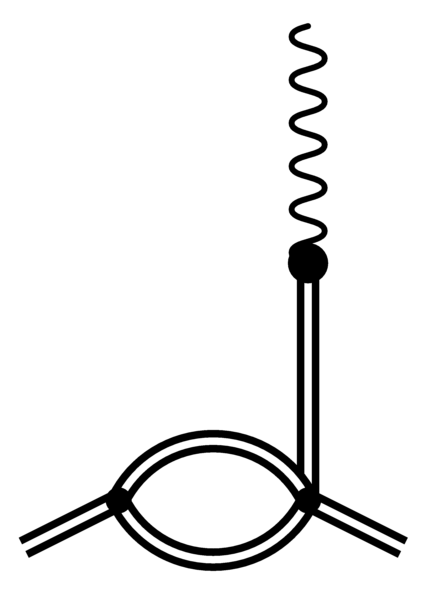}{fig:Loop11}
		\myfig{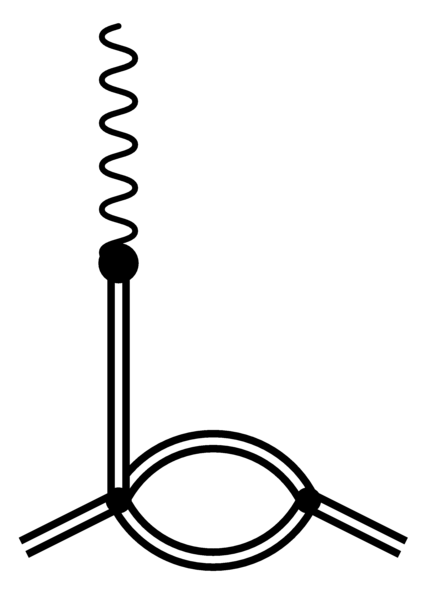}{fig:Loop10}
		\myfig{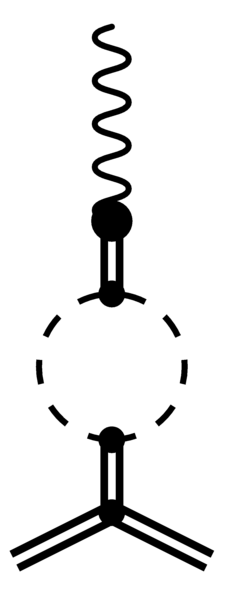}{fig:Loop12}
		\myfig{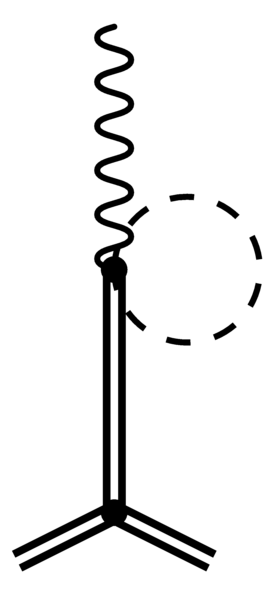}{fig:Loop13}
		\myfig{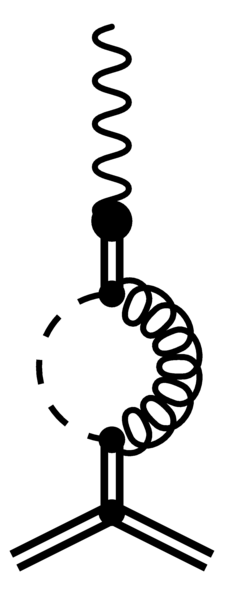}{fig:Loop14}
		\myfig{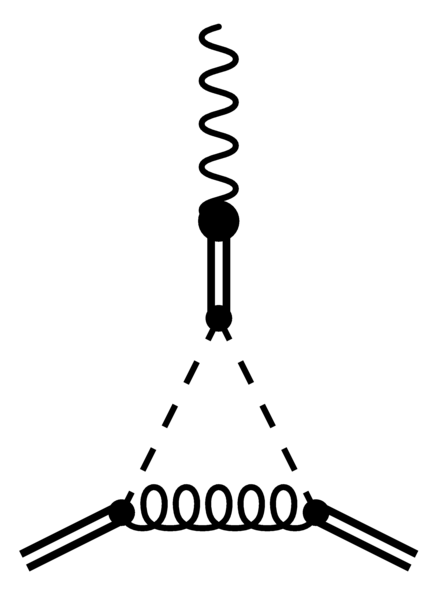}{fig:Loop15}
		\myfig{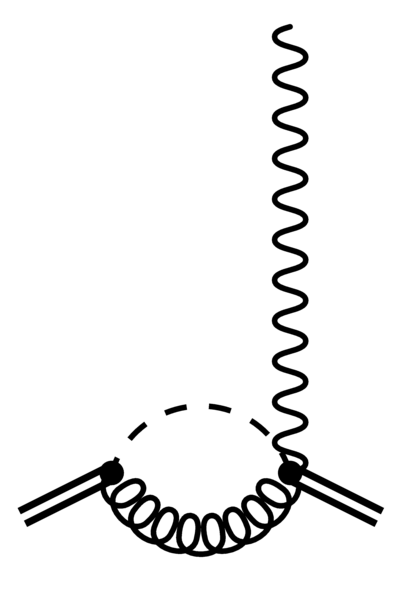}{fig:Loop17}
		\myfig{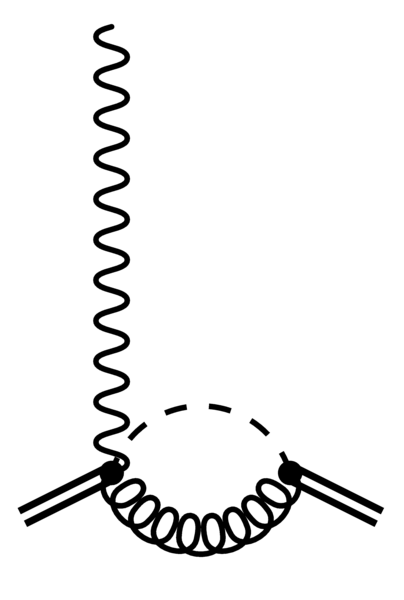}{fig:Loop16}
		\myfig{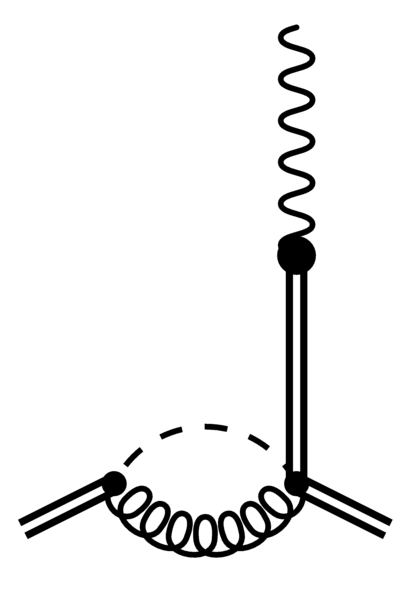}{fig:Loop19}
		\myfig{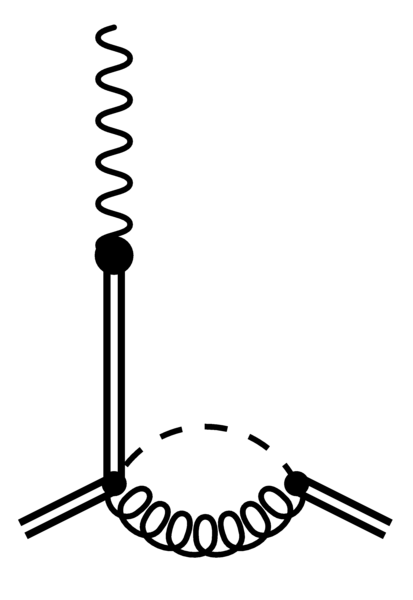}{fig:Loop18}
		\myfig{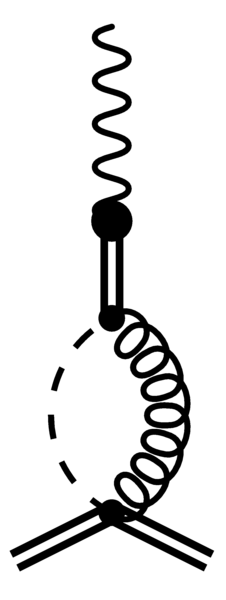}{fig:Loop20}
		\myfig{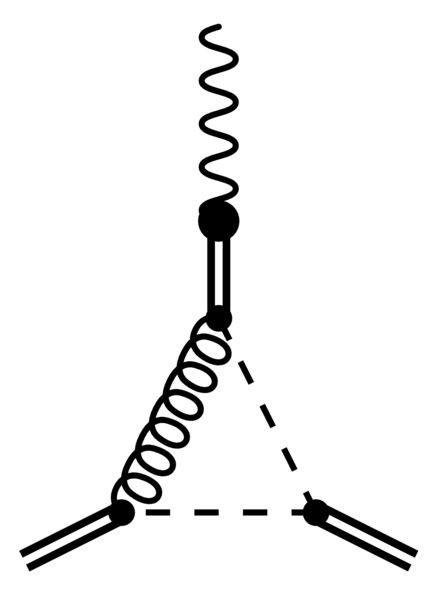}{fig:Loop21}
		\myfig{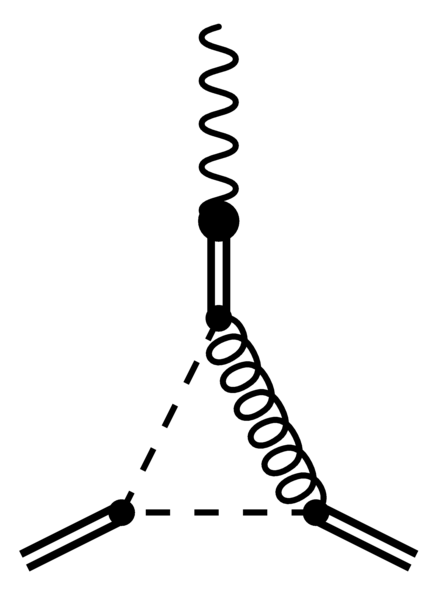}{fig:Loop22}
		\myfig{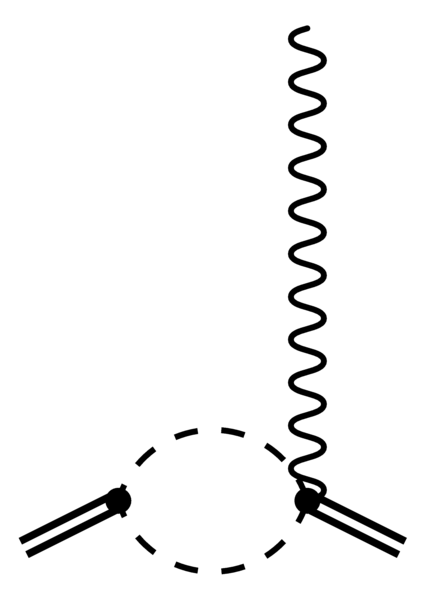}{fig:Loop24}
		\myfig{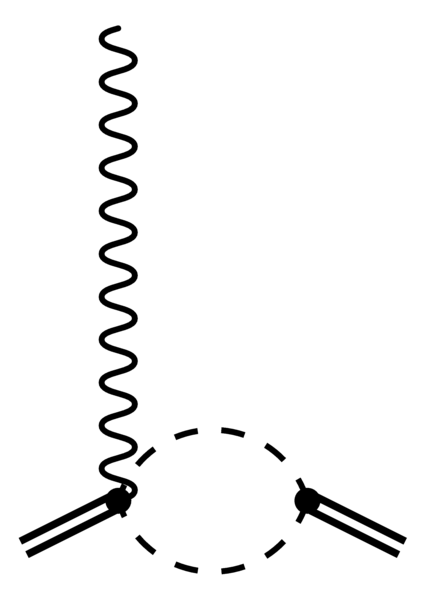}{fig:Loop23}
		\caption{Tree and one-loop diagrams contributing to the $\rho$-meson form factors.
			Wavy, squiggly, dashed and  double lines correspond to the photon, $\omega$,
			$\pi$ and $\rho$-mesons, respectively. Direct $\rho$-meson-photon
			couplings are shown by diagram~(2). }
		\label{fig:Diagrams}
	\end{figure}

	At the tree level, only the diagrams (1) and (2) shown in Fig.~\ref{fig:Diagrams} contribute
	to the invariant form factors, and the results read
\eq\label{eq:f1tree}
	f_1^{\textrm{tree}}(q^2)&=&-e\left[ \frac{(m_{\rho,0}^2+c_{x,0} M^2-g_0 D_x q^2)}{q^2-(m_{\rho,0}^2+c_{x,0}M^2)}
	+\frac{ h_V}{m_{\rho,0}^2}q^2\right],\\[2mm]
	\label{eq:f2tree}f_2^{\textrm{tree}}(q^2)&=&-e\left[\frac{2(m_{\rho,0}^2+c_{x,0}M^2-g_0 D_x q^2)}{q^2-(m_{\rho,0}^2
		+c_{x,0}M^2)}+ g_0 D_x+\frac{2 s_R}{m_{\rho,0}^2} h_V\right], \\[2mm]
	\label{eq:f3tree}f_3^{\textrm{tree}}(q^2)&=&-\, e \frac{4 s_R}{m_{\rho,0}^2} h_V \,,
	\en
	where $D_x=d_x-\sqrt{2}f_V$. After canceling polynomial contributions coming from
    the one-loop diagrams, $m_{\rho,0}^2,c_{x,0},g_0$ get replaced by $m_\rho^2,c_x,g$, respectively.
	
	We list all the one-loop diagrams (3)-(22) contributing to the form-factors also in
	Fig.~\ref{fig:Diagrams}. Importantly, we only consider the LECs $c_x$, $d_x$, $f_V$ and
	$h_V$ at tree-level, since their insertions in the loops generate diagrams beyond the
	accuracy of our calculation. For this reason, one-loop diagrams with vertices coming from
	either direct $\rho\rho\pi\pi$, $\gamma\pi\pi$, $\gamma\rho\rho$ or
	$\gamma\rho\rho\pi\pi$ couplings are not included. We use dimensional regularization
	to regulate divergences, carrying out calculations in the standard manner.
	Schematically, the results for the form factors can be written in the following form
	\eq
	f_i(q^2)=\sum_{\alpha=1}^{13}a_{i\alpha} I_\alpha, \qquad i=1,2,3\,,
	\en
	\newpage
	\noindent where $a_{i\alpha}$ are rational functions of $q^2$, masses and spacetime dimension $D$, and the $I_\alpha$ represent the following scalar integrals:
	\eq\label{eq:Ia}
	I_1  &=& C_0(m_\rho^2,m_\rho^2,q^2,M^2,M^2,M^2),\nonumber\\[2mm]
	I_2  &=& C_0(m_\rho^2,m_\rho^2,q^2,M^2,M^2,m_\omega^2),\nonumber\\[2mm]
	I_3  &=& C_0(m_\rho^2,m_\rho^2,q^2,M^2,m_\omega^2,M^2),\nonumber\\[2mm]
	I_4  &=& C_0(m_\rho^2,m_\rho^2,q^2,m_\rho^2,m_\rho^2,m_\rho^2),\nonumber\\[2mm]
	I_5  &=& B_0(m_\rho^2,m_\rho^2,m_\rho^2),\nonumber\\[2mm]
	I_6  &=& B_0(q^2,m_\rho^2,m_\rho^2),\nonumber\\[2mm]
	I_7  &=& B_0(q^2,M^2,M^2),\nonumber\\[2mm]
	I_8  &=& B_0(m_\rho^2,M^2,m_\omega^2),\nonumber\\[2mm]
	I_9  &=& B_0(q^2,M^2,m_\omega^2),\nonumber\\[2mm]
	I_{10} &=& B_0(m_\rho^2,M^2,M^2),\nonumber\\[2mm]
	I_{11} &=& A_0(M^2),\nonumber\\[2mm]
	I_{12} &=& A_0(m_\rho^2),\nonumber\\[2mm]
	I_{13} &=& A_0(m_\omega^2).
	\label{eq:integral-list}
	\en
	The loop functions $A_0$, $B_0$ and $C_0$ are defined as follows:
	\eq	\label{eq:ABC-definitions}
	&&\hspace{1.6cm}	A_0(m^2) = \frac{(2\pi)^{4-D}}{i\pi^2}\int \frac{d^D k}{k^2-m^2}~,\nonumber\\[2mm]
	&&\hspace{4mm}B_0(p^2,m_1^2,m_2^2) = \frac{(2\pi)^{4-D}}{i\pi^2}\int \frac{d^D k}{(k^2-m_1^2)((p+k)^2-m_2^2)}~,\nonumber\\[2mm]
	&&C_0(p_1^2,(p_2-p_1)^2,p_2^2,m_1^2,m_2^2,m_3^2)\nonumber\\
	&&\hspace{3cm}=\frac{(2\pi)^{4-D}}{i\pi^2}\int \frac{d^D k}{(k^2-m_1^2)((p_1+k)^2-m_2^2)((p_2+k)^2-m_3^2)}~.
	\en
	In the appendix~\ref{app:individual} we list the individual contributions of the
	one-loop diagrams, as well as give the full results in terms of scalar one-loop integrals.

	\section{Infrared regularization and counting rules}
	\label{sec:subtract}

	\subsection{Essentials}
	
	As already mentioned, the loop diagrams calculated  using the
	Lagrangian of Eq.~(\ref{eq:L-full}) break the power counting. The well-known reason for this
	is the presence of the heavy scale in the propagators associated with the
	$\rho$-  and $\omega$-meson masses. However, the power-counting-breaking terms are
	polynomials in external momenta and can be subtracted from the Green
	functions that amounts to the change of the renormalization prescription. In the literature
	one encounters different ways of setting such a systematic subtraction scheme
	in the presence of the massive vector mesons. 
	We mention here only two of them:  the complex mass renormalization 
	scheme~\cite{Bauer:2012at,Djukanovic:2009zn,Djukanovic:2015gna} and the infrared
	regularization of Refs.~\cite{Bruns:2004tj,Bruns:2008ub}. The scheme which we shall be using
	in this paper in fact represents a combination of the above two
	approaches. Let us stress here that subtraction schemes should 
	obey certain constraints. Namely,
	\begin{itemize}
		\item
		The $\rho$-meson in the Lagrangian from Sect.~\ref{sec:Lagrangian} is a gauge
		field of the hidden local symmetry. The subtractions should leave this symmetry
		intact. Note,  however, that the local hidden symmetry approach is just one 
		particular way, which is fully equivalent to the massive Yang-Mills approach, see
		Ref.~\cite{Meissner:1986tc} and the Ward identities in that approach are discussed in 
		Ref.~\cite{Gegelia:2011fp}.
		\item
		The electromagnetic current is conserved, and the electric form factor of the
		$\rho$-meson is normalized to unity at the origin by virtue of the $U(1)$ Ward identity.
		This Ward identity should not get upset by subtractions.
		\item
		It is very important to realize that the subtractions should work in different momentum
		regimes. For example, in order to renormalize the coupling $g$, or the $\rho$-meson
		mass, we should consider the diagrams with two, three,  and four external $\rho$-meson
		legs near the mass shell. In order to calculate the $\rho$-meson form  factors, however,
		we need the three-point function with two hard and one soft external momenta. It is not
		{\it a priori} clear, whether the power counting can be recovered in these
		two different momentum regions simultaneously.
		
	\end{itemize}
	
	Below, we shall address all these  issues.
	The procedure that we will be using,
	can be described as follows. At the first stage, we calculate the Green
	functions for {\em real} external momenta only. As usual,
	the self-energy insertions into the
	external lines corresponding to the $\rho$-mesons are summed up. Requiring real
	momenta then means that the external $\rho$-mesons are off-shell. However, their
	momenta are still counted as $O(1)$.
	
	At the next stage, we simplify the numerators in the Feynman diagrams. All these
	are reduced to the scalar integrals $I_\alpha,\,\alpha=1,\ldots, 13$ from Eq.~(\ref{eq:Ia}).
	The subtraction is applied to these scalar functions. To this end, we use infrared
	regularization, splitting off a polynomial $I_\alpha^R$ from $I_\alpha$
	\eq
	I_\alpha= I_\alpha^S+I_\alpha^R\, .
	\en
	By definition, $I_\alpha^R$, which is the regular part of $I_\alpha$, contains only integer
	powers of small kinematic variables in $D$ space-time dimensions up to a given
	chiral order.
	
	Finally, we continue the subtracted amplitudes to the second Riemann sheet
	and use  the complex mass scheme for the external $\rho$-mesons.
	Technical details concerning individual diagrams will be explained in the following
	sections.
	
	\subsection{Individual integrals}
	
	\subsubsection{The integral $I_1$}
	
	Using the Feynman parameterization, $I_1$ can be written as 
	\eq\label{eq:G1}
	I_1 &=& -\frac{\Gamma(3-D/2)}{(4\pi)^{D/2-2}}\,\int_0^1 xdx\int_0^1dy
	G_1(x,y)^{D/2-3}\, ,
	\nonumber\\[2mm]
	G_1(x,y) & = & A_1+B_1x^2-x(1-x)-i\varepsilon\, ,
	\nonumber\\[2mm]
	A_1 & = & M^2=O(q^2)\, ,
	\quad\quad
	B_1=-q^2\,y(1-y)=O(q^2)\, .
	\en
	Here, in conformity with Ref.~\cite{Becher:1999he}, we have fixed the renormalization scale $\mu$
	equal to $m_\rho$\footnote{This was earlier stressed in Ref.~\cite{Bernard:1992qa}.} and, furthermore, in order to ease the 
	notation, we use $m_\rho=1$
	everywhere (the full result can be easily restored by using dimensional arguments). Performing integration
	over the variable $x$, we arrive at the following result
	\eq\label{eq:F1}
	I_1&=&-\int_0^1 dy \frac{F_1(y)}{1+B_1}\, ,
	\nonumber\\[2mm]
	F_1(y)&=&\frac{1+\sigma_1}{2\sigma_1}\,\ln(M^2-q^2y(1-y)-i\varepsilon)
	+\frac{1-\sigma_1}{2\sigma_1}\,\ln M^2+\frac{\pi i}{\sigma_1}
	\nonumber\\[2mm]
	&+&\frac{1}{\sigma_1}\,\left(\ln(4(1+B_1))-\ln(1+2B_1+\sigma_1)-\ln\sigma_1)\right)\, ,
	\nonumber\\[2mm]
	\sigma_1&=&\sqrt{1-4A_1(1+B_1)}\, .
	\en
	One can now carry out the remaining integration in the variable $y$.
	To this end, one first performs chiral expansion in $F_1(y)$, taking into account
	that $\sigma_1=1+O(q^2)$. The calculations are pretty straightforward, and we
	quote the final result only
	\eq
	I_1
	&=&-
	\left(1+2M^2+\frac{q^2}{6}+6M^4+\frac{q^4}{30}\right)\ln M^2
	\nonumber\\[2mm]
	&-&
	\frac{\sqrt{4M^2-q^2}}{15\sqrt{-q^2}}\,
	\left(30+40M^2+5q^2+96M^4+2q^2M^2+q^4\right)\mbox{arctanh}\frac{\sqrt{-q^2}}
	{\sqrt{4M^2-q^2}}
	\nonumber\\[2mm]
	&-&
	i\pi\left(1+2M^2+\frac{q^2}{6}+6M^4+\frac{q^4}{30}\right)
	\nonumber\\[2mm]
	&-&
	\left(-2-\frac{2M^2}{3}-\frac{5q^2}{18}+\frac{3M^4}{5}+\frac{M^2q^2}{15}-\frac{47q^4}{900}\right)+O(q^6)\, .
	\en
	Next, we calculate the singular part of the same integral by using the infrared regularization. 
	To this end, we rewrite the function $G_1(x,y)$, appearing in Eq.~(\ref{eq:G1}), as
	\eq
	G_1(x,y)=(1+B_1)(x-x_1)(x-x_2)\, ,\quad\quad x_{1,2}=\frac{1\pm\sigma_1}{2(1+B_1)}\, .
	\en
	The integral over the variable $x$ can be transformed as
	\eq
	&&\int_0^1 xdx\,G_1(x,y)^{D/2-3}
	\nonumber\\[2mm]
	&& \qquad =(1+B_1)^{D/2-3}
	\int^{\frac{1+2B_1}{2(1+B_1)}}_{-\frac{1}{2(1+B_1)}}
	\left(x+\frac{1}{2(1+B_1)}\right)\left(x^2-\frac{\sigma_1^2}{4(1+B_1)^2}\right)^{D/2-3}
	\nonumber\\[2mm]
	&& \qquad =2^{3-D/2}(2(1+B_1))^{1-D/2}\int_{-1}^{1+2B_1}dx(1+x)(x^2-\sigma_1^2)^{D/2-3}
	\nonumber\\[2mm]
	&& \qquad =2^{3-D/2}(2(1+B_1))^{1-D/2}(J_1+J_1')\, .
	\en 
	Here, we have split the integration interval into $x\in [-1,1]$ and $x\in [1,1+2B_1]$. The first
	integral is transformed in the following manner
	\eq
	J_1&=&\int_{-1}^1dx(1+x)(x^2-\sigma_1^2)^{D/2-3}=2\int_0^1 dx(x^2-\sigma_1^2)^{D/2-3}
	\nonumber\\[2mm]
	&=&\int_0^1\frac{du}{\sqrt{u}}(u-1+4A_1(1+B_1))^{D/2-3}
	\nonumber\\[2mm]
	&=&\int_0^1\frac{du}{\sqrt{1-u}}(4A_1(1+B_1)-u)^{D/2-3}\, .
	\en
	In order to single out the singular part of this integral, we extend the integration over
	the Feynman parameter $u$ from $0$ to infinity, rescale $u\to 4A_1(1+B_1)u$ and expand the square root in the integrand. 
	In short, we get non-integer powers of the small variable $4A_1(1+B_1)$
	\eq
	J_1=\left(4A_1(1+B_1))^{D/2-2}(k_0+k_1(4A_1(1+B_1))+\cdots\right)\, .
	\en
	It is also easy to see that the ``dropped'' terms can be expanded in the integer powers of
	the same variable.
	
	The second integral can be treated similarly
	\eq
	J_1'&=&\int_1^{1+2B_1}dx(1+x)(x^2-\sigma_1^2)^{D/2-3}
	\nonumber\\[2mm]
	&=&
	\int_1^{(1+2B_1)^2}\frac{du }{2\sqrt{u}}\,(1+\sqrt{u})(u-1+4A_1(1+B_1))^{D/2-3}
	\nonumber\\[2mm]
	&=&\int_{-4B_1(1+B_1)}^0\frac{du}{\sqrt{1-u}}\,(1+\sqrt{1-u})(4A_1(1+B_1)-u)^{D/2-3}\, .
	\en
	As in the previous case, the non-integer powers can be singled out by expanding the
	integrand in powers of $u$ and integrating term by term. Here, there is no need to
	extend the integration interval to infinity.
	
	Putting all pieces together, we can write down the expression for the singular part
	of the integral $I_1$
	\eq
	I_1^S
	&=&
	-\left(1+2M^2+q^2/6+6M^4+q^4/30\right)\ln M^2
	\nonumber\\[2mm]
	&-&
	\frac{\sqrt{4M^2-q^2}}{15\sqrt{-q^2}}\,
	\left(30+40M^2+5q^2+96M^4+2q^2M^2+q^4\right)\mbox{arctanh}\frac{\sqrt{-q^2}}
	{\sqrt{4M^2-q^2}}
	\nonumber\\[2mm]
	&-&
	16\pi^2 \hat L\left(2+4M^2+\frac{q^2}{3}+12M^4+\frac{q^4}{15}\right)
	\nonumber\\[2mm]
	&-&
	\left(-2-\frac{14M^2}{3}-\frac{q^2}{9}-\frac{77M^4}{5}+\frac{2M^2q^2}{5}-\frac{q^4}{450}\right)+O(q^6)\, ,
	\en
	where $\hat L$ denotes\footnote{We remind the reader that the renormalization
		scale is chosen as $\mu=m_\rho$. Note also that this expression differs from the one
		from the standard convention used in ChPT by a constant term $+1$. We stick to this definition throughout this paper.}
	\eq
	\hat L=\frac{m_\rho^{D-4}}{16\pi^2}\,\biggl(\frac{1}{D-4}-\frac{1}{2}\,(\Gamma'(1)
	+\ln 4\pi)\biggr)\, .
	\en
	Finally, the regular part $I_1^R$ can be obtained, subtracting $I_1^S$ from $I_1$. The
	non-analytic terms cancel (as they should), and the result looks as follows
	\eq
	I_1^R
	&=&
	-i\pi\left(1+2M^2+\frac{q^2}{6}+6M^4+\frac{q^4}{30}\right)
	\nonumber\\[2mm]
	&+&
	\hat L\left(2+4M^2+\frac{q^2}{3}+12M^4+\frac{q^4}{15}\right)
	\nonumber\\[2mm]
	&-&\left(4M^2-\frac{q^2}{6}+16M^4-\frac{M^2q^2}{3}-\frac{q^4}{20}\right)+O(q^6)\, .
	\en
	
	\subsubsection{The integral $I_2$}
	
	This integral can be written in the form
	\eq\label{eq:G2}
	I_2&=&-\frac{\Gamma(3-D/2)}{(4\pi)^{D/2-2}}\,\int_0^1 xdx\int_0^1dy
	G_2(x,y)^{D/2-3}\, ,
	\nonumber\\[2mm]
	G_2(x,y)&=&m_\omega^2x(1-y)+M^2(1-x+xy)-x(1-x)-x^2y(1-y)q^2-i\varepsilon\, .
	\en
	As in the case of $I_1$, first we evaluate this integral for $D=4$, where it can be written as
	\eq\label{eq:I2}
	I_2=-\frac{2}{q^2}\,\int_0^1xdx F_2(x)\, ,
	\en
	where
	\eq
	F_2(x)&=&\int_{-1}^1 \frac{dy}{(xy+A_2)^2-B_2^2-i\varepsilon}\, ,
	\nonumber\\[2mm]
	A_2&=&-\frac{m_\omega^2-M^2}{q^2}\, ,\quad\quad
	B_2^2=A_2^2(1+C_2) \,,
	\nonumber\\[2mm]
	C_2&=&-\frac{2xq^2}{m_\omega^2-M^2}-\frac{4M^2q^2}{(m_\omega^2-M^2)^2}
	+\frac{4x(1-x)q^2}{(m_\omega^2-M^2)^2}+\frac{x^2q^4}{(m_\omega^2-M^2)^2}\, .
	\en
	Note that, in order to arrive at this expression, we used the substitution $y\to \frac{1}{2}\,(1+y)$.
	Carrying out the integration in the variable $y$, we get
	\eq\label{eq:F2}
	F_2(x)=-\frac{q^2}{2x(m_\omega^2-M^2)\sqrt{1+C_2}}\,
	\biggl(
	\ln\frac{(x+A_2)^2-B_2^2-i\varepsilon}{(-x+A_2)^2-B_2^2-i\varepsilon}
	-2\ln\frac{x+A_2+B_2}{-x+A_2+B_2}\biggr)\, .\quad\quad
	\en
	Furthermore, $A_2=B_2=O(q^{-2})$ and $C_2=O(q^2)$. One may safely perform the
	chiral expansion of the prefactor in Eq.~(\ref{eq:F2}), as well as of the second logarithm
	in the brackets. No singularities ever appear. The argument of the first logarithm
	is given by
	\eq
	\frac{(x+A_2)^2-B_2^2-i\varepsilon}{(-x+A_2)^2-B_2^2-i\varepsilon}
	=\frac{M^2-x(1-x)}{x^2+M^2+x(\Delta-M^2)}\, ,
	\en
	and it cannot be expanded due to the resulting divergent integrals. Performing the expansion and integrating over the variable $x$
	term by term, we finally get
	\eq
	I_2
	&=&
	i\pi\,\left(-1+\left(M^2-\frac{q^2}{6}+\Delta\right)
	+\left(3M^4-q^2M^2-\frac{q^4}{30}-\Delta^2\right)\right)
	\nonumber\\[2mm]
	&+&\ln(M^2)\left(\left(\frac{3M^2}{2}+\frac{\Delta}{2}\right)
	+\left(\frac{7M^4}{2}+\frac{q^2M^2}{2}-\Delta M^2-\frac{\Delta^2}{2}\right)\right)
	\nonumber\\[2mm]
	&+&H_2\left(-2M^2-\left(\frac{3M^4}{2}-\Delta M^2-\frac{\Delta^2}{2}\right)\right)
	\nonumber\\[2mm]
	&+&\left(\left(-2M^2+\frac{q^2}{2}-\Delta\right)+\left(\Delta^2+\frac{3M^2\Delta}{2}
	-q^2\Delta-M^4+\frac{M^2q^2}{2}+\frac{q^4}{4}\right)\right)\, {\color{red} ,}
	\en
	where
	\eq\label{eq:H2}
	H_2&=&\frac{1}{\beta}\left(\frac{\pi}{2}-\arctan\frac{\beta}{1-\alpha}+\arctan\frac{\alpha}{\beta}\right)\, ,
	\nonumber\\[2mm]
	\alpha&=&\frac{1}{2}\,(M^2-\Delta)\, ,\quad\quad \beta^2=M^2-\frac{1}{4}(\Delta-M^2)^2\, .
	\en
	In order to calculate the singular part of the integral, we go back to the expression in
	Eq.~(\ref{eq:G2}), written in $D$ dimensions. The function $G_2(x,y)$ after
	the change of variable $y\to \frac{1}{2}\,(1+y)$ can be written as
	\eq
	G_2(x,y)=\frac{q^2}{2}\,A_2\biggl(1+\frac{xy}{2A_2}+\frac{B_2-A_2}{2A_2}\biggr)
	(xy+A_2-B_2-i\varepsilon)\, .
	\en
	Substituting this in Eq.~(\ref{eq:G2}) and taking into account that $A_2=O(q^{-2})$,
	$B_2-A_2=O(1)$, it is seen that the expression
	$\left(1+\dfrac{xy}{2A_2}+\dfrac{B_2-A_2}{2A_2}\right)^{D/2-3}$ can be chirally
	expanded. The integration over the variable $y$ can be done term by term, and the
	result  is:
	\eq
	I_2&=&-\frac{\Gamma(3-D/2)}{2(4\pi)^{D/2-2}}\,
	\left(\frac{q^2A_2}{2}\right)^{D/2-3}
	\nonumber\\[2mm]
	&\times&\int_0^1xdx\biggl(J_2^{(+)}(x)(x+A_2-B_2)^{D/2-2}-J_2^{(-)}(x)(-x+A_2-B_2)^{D/2-2}\biggr)\, ,
	\en
	where $J_2^{(\pm)}(x)$ can be expanded in series of  $x$. Next, using the identities
	\eq
	x+A_2-B_2&=&\frac{2(M^2-x(1-x))}
	{A_2\left(1+\dfrac{x}{2A_2}+\dfrac{B_2-A_2}{2A_2}\right)}\, ,
	\nonumber\\[2mm]
	-x+A_2-B_2&=&\frac{2(x^2+M^2+x(\Delta-M^2))}
	{A_2\left(1-\dfrac{x}{2A_2}+\dfrac{B_2-A_2}{2A_2}\right)}\, ,
	\en
	one may rewrite the integral $I_2$ in the following form:
	\eq
	I=\frac{\Gamma(3-D/2)}{(4\pi)^{D/2-2}}\,\frac{1}{1+ \Delta-M^2}\,\left(I_2^{(+)}-I_2^{(-)}\right)\, ,
	\en
	where
	\eq
	I_2^{(+)}&=&\int_0^1dx\,\tilde J_2^{(+)}(x)(M^2-x(1-x))^{D/2-2}\, ,
	\nonumber\\[2mm]
	I_2^{(-)}&=&\int_0^1dx\,\tilde J_2^{(-)}(x)(x^2+x(\Delta-M^2)+M^2)^{D/2-2}\, .
	\en
	The quantities $\tilde J_2^{(\pm)}(x)$ are given by the products of $J_2^{(\pm)}(x)$ and
	the factors that can be chirally expanded. At any order  of this expansion, the resulting expressions are polynomials in $x$.

	At this stage, we are ready to extract the singular piece from the integral $I_2$.
	To this end, consider first $I_2^{(+)}$. Making the substitution $x\to \frac{1}{2}\,(1+x)$,
	we get
	\eq
	I_2^{(+)}=\frac{1}{2}\,\int_{-1}^1\tilde J_2^{(+)}(x)\left(M^2-\frac{1}{4}\,(1-x^2)\right)^{D/2-2}\, .
	\en
	As already mentioned, $\tilde J_2^{(+)}$ can be expanded in $x$
	as $\tilde J_2^{(+)}=\sum\limits_{n=0}^\infty \tilde J_{2,n}^{(+)}x^n$. All odd powers
	give zero contribution and, hence,
	\eq
	I_2^{(+)}&=&\sum_{n=0}^\infty\tilde J_{2,n}^{(+)}\frac{1}{2}\,
	\int _{-1}^1  dx\,x^{2n}\left(M^2-\frac{1}{4}\,(1-x^2)\right)^{D/2-2}
	\nonumber\\[2mm]
	&=&\sum_{n=0}^\infty\tilde J_{2,n}^{(+)}2^{3-D}
	\int_0^{1}\int_0^1du\,(1-u)^{n-1/2}(4M^2-u)^{D/2-2}\, .
	\en
	In order to single out the singular part, one should extend the integration range
	in the variable $u$ from $[0,1]$ to $[0,\infty[$, and expand the square root. In essence, 
	only non-integer powers will survive.
	
	Turning to the integral $I_2^{(-)}$, one can first transform it into the form 
	\eq
	I_2^{(-)}=\int_0^1dx J_2^{(-)}(x)\left((x-\alpha)^2+\beta^2\right)^{D/2-2}\, ,
	\en
	where $\alpha,\beta$ are defined in Eq.~(\ref{eq:H2}). Here, in order
	to single out the singular piece, the integration in $x$ is extended to infinity and the rescaling $x\to\beta x$ is performed. One finds 
	\eq
	I_2^{(-)}&=&\beta^{D-3}\int_0^\infty dx\, J_2^{(-)}(\beta x)\left(x^2+1-\frac{2\alpha}{\beta}\,x
	+\frac{\alpha^2}{\beta^2}\right)^{D/2-2}
	\nonumber\\[2mm]
	&=&\beta^{D-3}\int_0^\infty dx\, J_2^{(-)}(\beta x)(x^2+1)^{D/2-2}
	\left(1+\left(\frac{D}{2}-2\right)\frac{-\dfrac{2\alpha}{\beta}\,x
		+\dfrac{\alpha^2}{\beta^2}}{x^2+1}+\cdots\right)\, .\quad\quad
	\en
	Expanding  $J_2^{(-)}(\beta x)$ in powers of $\beta x$ and integrating term by term,
	we again arrive at the expression that contains only non-integer powers.
	
	Finally, putting all pieces together, we get the expression for the singular piece  $I_2^S$:
	\eq\label{eq:I2S}
	I_2^S
	&=&+\hat L\left(\left(3M^2+\Delta\right)
	-\left(7M^4+q^2M^2-2M^2\Delta-\Delta^2\right)\right)
	\nonumber\\[2mm]
	&+&\ln(M^2)\left(\left(\frac{3M^2}{2}+\frac{\Delta}{2}\right)
	+\left(\frac{7M^4}{2}+\frac{q^2M^2}{2}-\Delta M^2-\frac{\Delta^2}{2}\right)\right)
	\nonumber\\[2mm]
	&+&\pi M\left(-1-\frac{7M^2}{8}+\frac{\Delta^2}{8M^2}+\frac{3\Delta}{4}\right)
	\nonumber\\[2mm]
	&+&\left(-2M^2+\left(-\frac{\Delta^3}{12M^2}+\frac{\Delta^2}{4}+\frac{7\Delta M^2}{4}-\frac{59M^4}{12}-\frac{M^2q^2}{2}\right)\right)+O(q^6)\, ,
	\en
	and the regular part, $I_2^R=I_2-I_2^S$ is given by
	\eq
	I_2^R
	&=&i\pi\,\left(-1+\left(M^2-\frac{q^2}{6}+\Delta\right)
	+\left(3M^4-q^2M^2-\frac{q^4}{30}-\Delta^2\right)\right)
	\nonumber\\[2mm]
	&-&\hat L\left(\left(3M^2+\Delta\right)+\left(7M^4+q^2M^2-2M^2\Delta-\Delta^2\right)\right)
	\nonumber\\[2mm]
	&+&\left(\left(M^2+\frac{q^2}{2}\right)
	+\left(-\frac{\Delta^2}{2}-\frac{\Delta M^2}{2}+\frac{29 M^4}{6}
	+M^2q^2-q^2\Delta+\frac{q^4}{4}\right)\right)+O(q^6)\, .\quad\quad
	\en
	Note that the non-analytic piece, $\dfrac{\Delta^3}{M^2}$ in Eq.~(\ref{eq:I2S}) is
	canceled in the difference, as it should.
	
	\subsubsection{The integral $I_{10}$}
	
	The expression for this quantity in $D$ dimension is given by
	\eq
	I_{10}=\frac{\Gamma(2-D/2)}{(4\pi)^{D/2-2}}\,2^{4-D}\int_0^1dx(x^2-\sigma^2)^{D/2-2}\, ,\quad\quad
	\sigma^2=1-4M^2\, .
	\en
	Expanding this expression in chiral powers, we get
	\eq
	I_{10}  &=&-2\hat L+i\pi(1-2M^2-2M^4)
	-\ln(M^2)(2M^2+2M^4)
	\nonumber\\[2mm]
	&+&(2+2M^2-M^4)+O(q^6)\, .
	\en
	In order to calculate the singular part, we transform this integral as
	\eq
	I_{10}=\frac{2\Gamma(2-D/2)}{(4\pi)^{D/2-2}}\,2^{3-D}
	\int_0^1\frac{du}{\sqrt{1-u}}\,(4M^2-u-i\varepsilon)^{D/2-2}\, .
	\en
	Next, we extend the integration in the variable $u$ to infinity, expand the square root
	and  calculate the integral term by term. In $D=4$ dimensions we then obtain
	\eq
	I_{10}^S
	=-2\hat L(2M^2+2M^4)-\ln(M^2)(2M^2+2M^4)+(2M^2+3M^4)+O(q^6)\, .
	\en
	Finally, the regular part is obtained by subtracting these two expressions
	\eq
	I_{10}^R
	=-2\hat L(1-2M^2-2M^4)+i\pi(1-2M^2-2M^4)+2(1-2M^4)\, .
	\en
	Note that, at leading order, this result was already obtained in Ref.~\cite{Bruns:2004tj}.
	
	\subsubsection{Other integrals}
	
	We do not consider the calculation of other integrals in detail, because this has already
	either been addressed in the literature, or is trivial.
	
	\begin{itemize}
		
		\item The integrals $I_3,I_8$ can be calculated by straightforwardly applying the
		prescription of Ref.~\cite{Becher:1999he}.
		
		\item It can be easily checked that this prescription also works for $I_9$. Indeed, it is given by the expression
		\eq
		I_9=\frac{\Gamma(2-D/2)}{(4\pi)^{D/2-2}}\,\int_0^1dx\,
		(xm_\omega^2+(1-x)M^2-x(1-x)q^2)^{D/2-2}\, .
		\en
		It is immediately seen that the non-integer powers emerge from the vicinity of $x=0$.
		The regular part is given by the same integral with minus sign, taken from 1 to infinity.
		\item
		The integrals $I_4,I_5,I_6,I_{12},I_{13}$ do not contain singular pieces and are
		low-energy polynomials. They can be obtained by expanding the full expression up  to given chiral order.
		
		\item
		The integrals $I_7,I_{11}$ do not contain heavy scale and, therefore, their regular
		parts vanish.

	\end{itemize}
	
	In appendix~\ref{app:regular} we list the regular parts of all 13 integrals.\footnote{These expressions of infrared regular parts have been also verified by using the method of Ref.~\cite{Schindler:2003xv}.} 
	In the final expression for the amplitude, all $I_\alpha$ are replaced by $I_\alpha-I_\alpha^R$. 
	
	\subsection{The Ward identities} 
	
	Ward identities, which encode  the full symmetry content of the theory, represent a tower
	of linear relations between different Green functions. In its turn, these Green functions
	are decomposed into scalar integrals of the type of $I_\alpha$ considered above, with the
	coefficients that depend on external momenta and masses. These coefficients have integer
	mass dimension for any $D$. Thus, splitting the scalar integrals into the singular and regular
	pieces corresponds to splitting of the Green functions. Invoking the standard argument, one
	may conclude that the Ward identities should hold separately for the regular and singular
	parts, and that the subtraction does not destroy the symmetries. Note that below
	we shall explicitly check the fulfillment of the Ward identity associated with QED, which
	ensures the proper normalization of the electric form factor at $q^2=0$.
	
	The subtle issue that still remains is the fact that, dropping the regular parts of all $I_\alpha$,
	one rectifies the counting rules in two different regimes. In order to explain the difficulty,
	consider the one-loop corrections to the triple-$\rho$ vertex. It is clear that, in order to
	preserve the formulation based on hidden local symmetry, all three external legs with
	rho-mesons should be treated on equal footing (this must hold in any other representation
	of the vector mesons, too).
	For example, the counting rules
	should hold, when all external momenta in the diagram are considered as $O(1)$.
	Furthermore, one may attach an external photon to one of the outgoing $\rho$-meson
	lines via the direct coupling. This diagram will contribute to the electromagnetic form  factors
	for the small values of $q^2$. On the other hand, since dropping the regular part amounts
	to changing the renormalization prescription, it is legitimate to ask, whether there are
	necessary counterterms present in the Lagrangian that can absorb the change of the
	renormalization prescription for two different configurations of external momenta.
	
	After thoroughly examining the structure of the effective Lagrangian displayed in
	Sect.~\ref{sec:Lagrangian}, one concludes that the answer to the above question is
	{\em yes.} Indeed, the subtractions in the regime when all external momenta are large,
	can be associated with the operator $\mbox{Tr}(\rho_{\mu\nu}\rho^{\mu\nu})$, or with similar terms, 
	containing higher-order covariant derivatives (only gauge-invariant operators
	suffice because of  the Ward identities). The diagrams with the external
	electromagnetic field carrying small momentum are additionally renormalized by the
	contributions coming from the operators proportional to $d_x,f_V$ and $h_V$. These
	just suffice to reenact counting rules for small $q^2$ without destroying them for the
	triple-$\rho$ vertex. By the way, this discussion clearly
	shows that it is inconsistent to leave these
	operators out when calculating the electromagnetic form  factors of the $\rho$-meson.
	
	Last but not least, the subtraction polynomial is complex. Moreover, it is a polynomial
	with complex coefficients, so that the renormalized Lagrangian is not unitary. This
	does not, however, render the theory inconsistent. The imaginary parts of the couplings in
	the Lagrangian allow for a clear physical interpretation, they include the contributions
	from the distant intermediate states lying
	{\em below} the energy interval considered in the calculations. Note also that complex
	couplings do not automatically lead to the breaking of the time reversal invariance.
	In this case, only unitarity is broken (see, e.g.~\cite{Gasser:2007zt} for a detailed discussion).

	\subsection{Analytic continuation to the second sheet}
	
	In general, resonances correspond to the $S$-matrix poles on the unphysical sheets.
	In particular, the $\rho$-meson is located on the second Riemann sheet, which is connected to the physical sheet across the cut associated with the two-pion intermediate state. The cuts associated with the heavier intermediate states (say, the $\pi\omega$ intermediate state) define other sheets which are not considered here.
	
	The analytic continuation of the Feynman amplitudes to the unphysical Riemann sheets is a well-studied problem. A subtle  issue, which may still arise here is the interplay between the continuation and the subtraction of the regular part (which is a low-energy polynomial and does not have cuts at all). Namely, it can be observed that if the subtraction rectifies the counting rules on the physical sheet, it also rectifies them on the unphysical sheet which is directly connected to the physical one. Intuitively, this is crystal clear. Namely, the generic Feynman amplitude $f$ after subtraction is given in a form $f(z)=f_0(z)+f_1(z)+\cdots$, where $z$ denotes an (unspecified) kinematic variable, and $f_0,f_1,\ldots$ stand for the contributions at different chiral orders $f_0(z)\gg f_1(z)\ldots$. It is clear that changing $z$ slightly by going to the second sheet across the cut cannot unravel this hierarchy immediately, thus all changes should be continuous.
	
	As a rule, the discontinuity over
	the cut, arising in the one-loop diagrams is of a square-root type. Therefore, one may
	consider a prototype function as an example
	\eq
	f(p^2)=\sqrt{ \frac{4M^2}{p^2}-1}\, .
	\en
	In the upper half of the physical plane the variable $p^2$ has a positive imaginary part
	and the square root for  $\mbox{Re}\,p^2>4M^2$ can be rewritten as
	\eq
	f(p^2)=-i\sqrt{1-\frac{4M^2}{p^2}}\, .
	\en
	Exactly the same expression takes the square root in the lower half of the second sheet,
	now with $\mbox{Im}\,p^2<0$. In both cases, the chiral expansion gives
	\eq
	f(p^2)= -i \left( 1-\frac{2M^2}{p^2}+\cdots \right) \, , 
	\en
	and the extrapolation from $\mbox{Im}\,p^2>0$ to $\mbox{Im}\,p^2<0$ is smooth. Let us stress once more that this argument is immediately applicable to the narrow resonances on the sheets adjacent to physical sheet. However, if one has to move a long path from the physical sheet to reach the resonance in question, the argument is no more applicable. 
	
	After these preliminary comments we consider analytic continuation of the scalar
	integrals $I_\alpha$ and start again with $I_1$.
	For real values of the $p^2\doteq p_1^2=p_2^2$, and
	$q^2\leq 4M^2$, the imaginary part of the diagram is given by (see Eq.~(\ref{eq:F1}))
	\eq
	\delta_1\doteq\mbox{Im}\,I_1=-\pi\int_0^1\frac{dy}{\sigma_1(1+B_1)}\, .
	\en
	Here $\sigma_1,B_1$ are now defined for arbitrary values of $p^2$
	\eq
	A_1=\frac{M^2}{p^2}\, ,\quad\quad
	B_1=-\frac{q^2}{p^2}\,y(1-y)\, ,\quad\quad
	\sigma_1=\sqrt{1-4A_1(1+B_1)}\, .
	\en
	The integral on the second sheet is defined as
	\eq
	\tilde I_1=I_1-2i\delta_1\, ,
	\en
	evaluated now at arbitrary (complex) value of $p^2$.
	
	Next, let us consider the integral $I_2$ given by Eqs.~(\ref{eq:I2}) and (\ref{eq:F2}).
	 For the real values of $p^2$, the imaginary part of $I_2$ is given by
	\eq
	\mbox{Im}\,I_2=-\frac{1}{16\pi}\int^{\frac{1+\sigma}{2}}_{\frac{1-\sigma}{2}}
	\frac{dx}{(m_\omega^2-M^2)\sqrt{1+C_2}}\, ,\quad\quad
	\sigma=\sqrt{1-\frac{4M^2}{p^2}}\, .
	\en
	Substituting $x=\dfrac{1}{2}+\dfrac{\sigma}{2}\,z$, one obtains
	\eq
	\delta_2=\mbox{Im}I_2=-\frac{\sigma}{32\pi(m_\omega^2-M^2)}\,\int_{-1}^1\frac{dz}{\sqrt{1+C_2(z)}}\, ,
	\en
	and the integral $I_2$ on the second sheet is defined through $I_2\to I_2-2i\delta_2$
	($p^2$ is now complex).
	
	Finally, in case of $I_{10}$, we have $I_{10}=I_{10}-2i\delta_{10}$, where
	\eq
	\delta_{10}=\frac{1}{16\pi}\,\sqrt{1-\frac{4M^2}{p^2}}\, .
	\en
	Other integrals are not modified, only the argument is continued to the lower half-plane.

	\section{$U(1)$ Ward identities}
	\label{sec:Ward}
	
	The electric form factor of the $\rho$-meson, $f_1(q^2)$ in Eq.~(\ref{eq:Form Factors}),
	is normalized to unity at $q^2=0$ by virtue of the Ward identities. These $U(1)$ Ward
	identities boil down to certain linear relations between the scalar integrals $I_\alpha$ in
	$D$ dimensions. To derive these, we start with the identities
	\eq
	0&=&\int d^Dk\frac{\partial}{\partial k^\mu}\biggl(\frac{p^\mu}{(k^2-m_1^2)(k^2-m_2^2)}\biggr)
	\nonumber\\[2mm]
	&=&(m_1^2-m_2^2-p^2)I_{1,2}+(m_1^2-m_2^2+p^2)I_{2,1}-I_{2,0}+I_{0,2}\, ,
	\nonumber\\[2mm]
	0&=&\int d^Dk\frac{\partial}{\partial k^\mu}\biggl(\frac{k^\mu}{(k^2-m_1^2)(k^2-m_2^2)}\biggr)
	\nonumber\\[2mm]
	&=&(D-3)I_{1,1}-2m_1^2I_{1,1}-2m_1^2I_{2,1}-(m_1^2+m_2^2-p^2)I_{1,2}-I_{2,0}\, ,
	\en
	where
	\eq
	I_{1,2}&=&\int d^Dk\frac{1}{(k^2-m_1^2)((k+p)^2-m_2^2)^2}\, ,
	\nonumber\\[2mm]
	I_{2,1}&=&\int d^Dk\frac{1}{(k^2-m_1^2)^2((k+p)^2-m_2^2)}\, ,
	\nonumber\\[2mm]
	I_{1,1}&=&\int d^Dk\frac{1}{(k^2-m_1^2)((k+p)^2-m_2^2)}\, ,
	\nonumber\\[2mm]
	I_{2,0}&=&\int d^Dk\frac{1}{(k^2-m_1^2)^2}\, ,
	\nonumber\\[2mm]
	I_{0,2}&=&\int d^Dk\frac{1}{((k+p)^2-m_2^2)^2}\, .
	\en
	Furthermore, for the tadpole contributions, one easily obtains
	\eq
	I_{2,0}=\frac{D-2}{2m_1^2}\, I_{1,0}\, ,\quad\quad
	I_{0,2}=\frac{D-2}{2m_2^2}\, I_{0,1}\, .
	\en
	One can solve above equations for $I_{1,2},I_{2,1}$. In short, one gets
	\eq\label{eq:Ward-ini}
	I_{1,2}=\frac{2(3-D)m_2^2(m_1^2-m_2^2+p^2)I_{1,1}+(2-D)(m_1^2+m_2^2-p^2)I_{0,1}
		-2(2-D)m_2^2I_{1,0}}{2m_2^2(p^4-2p^2(m_1^2+m_2^2)+(m_1^2-m_2^2)^2)}\, .
	\nonumber\\
	\en
	The expression for $I_{2,1}$ is obtained from the above one
	via the substitution $m_1^2\leftrightarrow m_2^2$.
	
	Furthermore, since
	\eq
	I_{1,2}&=&\frac{i\pi^2}{(2\pi)^{4-D}}\,C_0(p^2,0,p^2,m_1^2,m_2^2,m_2^2)\, ,
	\nonumber\\[2mm]
	I_{1,1}&=&\frac{i\pi^2}{(2\pi)^{4-D}}\,B_0(p^2,m_1^2,m_2^2)\, ,
	\nonumber\\[2mm]
	I_{1,0}&=&\frac{i\pi^2}{(2\pi)^{4-D}}\,A_0(m_1^2)\, ,
	\nonumber\\[2mm]
	I_{0,1}&=&\frac{i\pi^2}{(2\pi)^{4-D}}\,A_0(m_2^2)\, ,
	\en
	the equation~(\ref{eq:Ward-ini}) can be rewritten  as a linear relation between different
	integrals $I_\alpha$. It is straightforward to check that $f_1(0)=1$, if these linear relations are obeyed.
	
	It remains to make sure that the subtractions do not upset the Ward identities. This
	fact should be self-evident, because non-analytic and analytic terms in these integrals
	obey the Ward identities separately. We have still carried out explicit checks
	for all possible cases: 1) $m_1^2=m_2^2=M^2$; 2) $m_1^2=m_\omega^2$,
	$m_2^2=M^2$; 3) $m_1^2=M^2$, $m_2^2=m_\omega^2$ and 
	4) $m_1^2=m_2^2=m_\rho^2$. All identities except the case 3) are explicitly fulfilled,
	whereas the case 3) is fulfilled only to the order one is working. Hence, the deviation of the
	calculated $f_1(0)$ from unity should be of higher order in chiral expansion.

	\section{Numerical results and discussion}\label{sec:numerics}

	We are now in a position to  calculate all three electromagnetic form factors.
	In these calculations we use the following input values
	$m^2_\rho+c_xM^2=0.775^2-i 0.775 \times 0.149\,\mbox{GeV}^2$,
	$m_\omega=0.782$~GeV, $M=0.13957$~GeV for the  masses,\footnote{
		In the loops, one may neglect the term proportional to $c_x$ in the $\rho$-meson mass.
		In addition, we neglect the width of the $\omega$-meson.}
	$F=F_\pi=0.0924$~GeV for the pion decay constant, and $g_{\omega\rho\pi}=1.478$ for the
	$\rho\omega\pi$ coupling.
	In the beginning of the discussion, the LECs $D_x$ and $h_V$ are set to zero.
	We shall see later, however, that their effect might be quite large.
	
\begin{figure}[t]	
	\myfigg{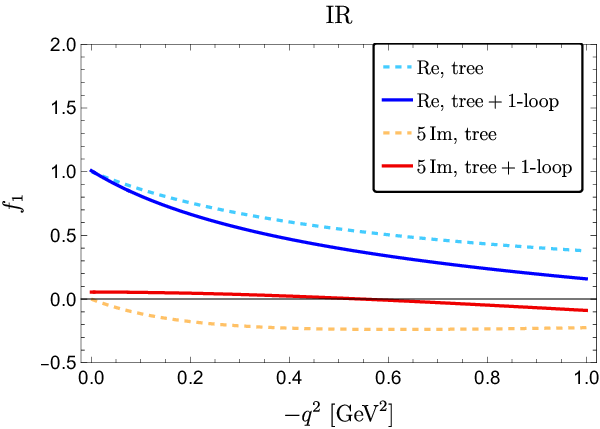}\hfill
	\myfigg{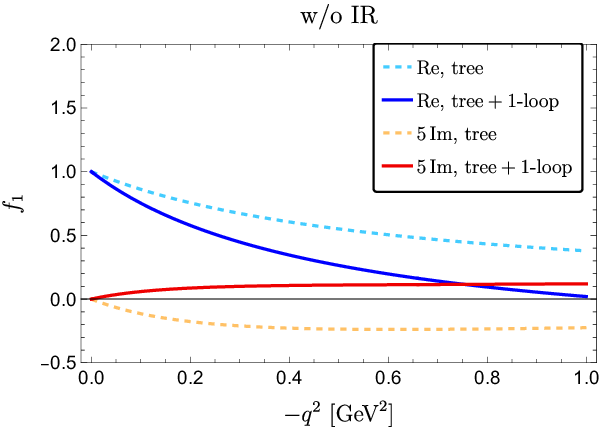}
	
	\vspace{5mm}	
	
	\myfigg{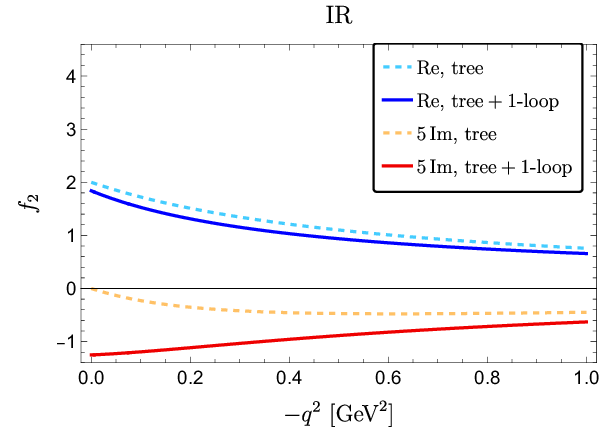}\hfill
	\myfigg{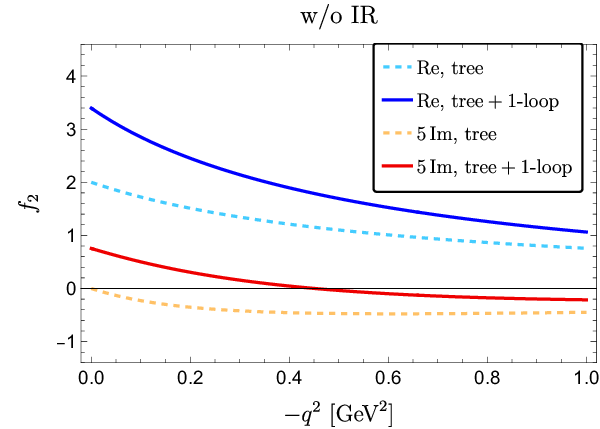}
	
	\vspace{5mm}
	
	\myfigg{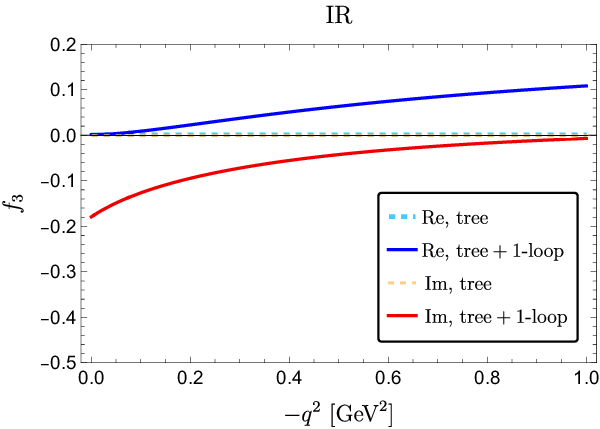}\hfill
	\myfigg{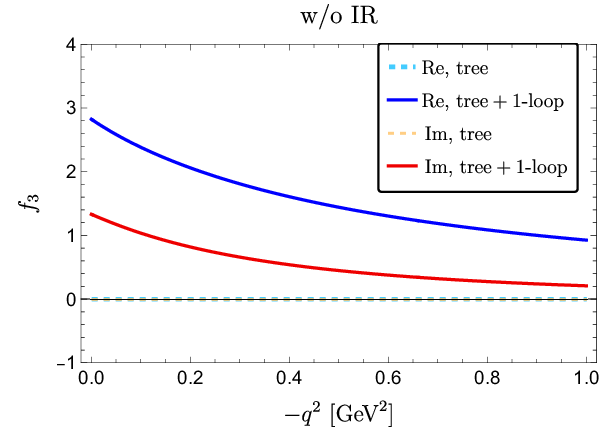}	
	
	\caption{Form factors $f_{1}(q^2),f_{2}(q^2),f_{3}(q^2)$ of the $\rho$ meson as functions of $q^2$ in the range $-1\,\mbox{GeV}^2<q^2<0$.
		The left column shows the results obtained with IR regularization, while the right column displays the results without IR regularization. Dashed curves denote the tree-level results, and solid curves represent the one-loop predictions. For better visibility, in the last row, the
		scale of the vertical axis was chosen differently on the left and right panels. }
	\label{fig:IRvsNoIR}
\end{figure}

	First, we would like to see, how  the infrared regularization affects the 
	convergence of the   perturbative series. To address this question, in
	Fig.~\ref{fig:IRvsNoIR} we plot the tree-level and the tree plus one-loop order results for
	each form factor. In the left column the subtractions are applied, whereas in the
	right column the unsubtracted results are displayed. It is seen that, as expected,
	for the real parts of $f_1(q^2)$ and $f_2(q^2)$ the subtraction of the regular pieces
	significantly improves the convergence. The improvement is less pronounced in
	the imaginary parts. For $f_3$, the situation is even more dramatic, both in the real
	and imaginary parts. Here, after subtraction, the one-loop result becomes very small,
	while the tree-level result from Eq.~(\ref{eq:f3tree}) is exactly zero.

\begin{figure}[t]
	\centering

	\myfiggg{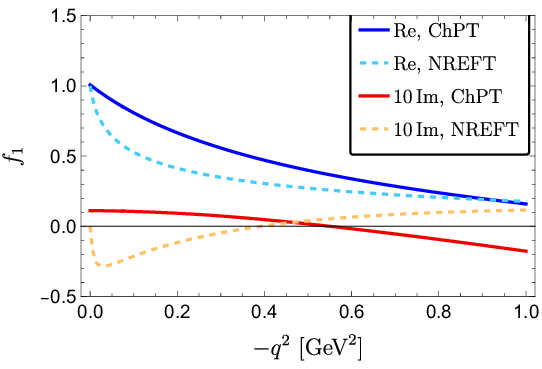}\hfill
	\myfiggg{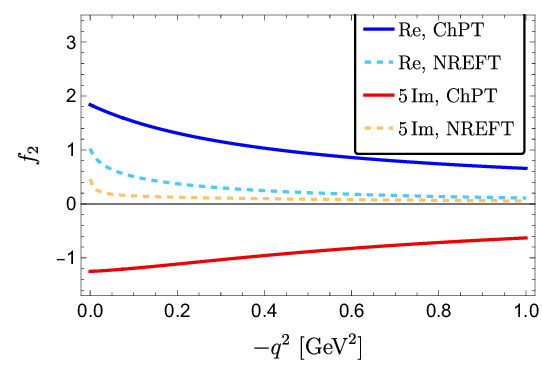}
	
	\vspace{5mm}
	
	\myfiggg{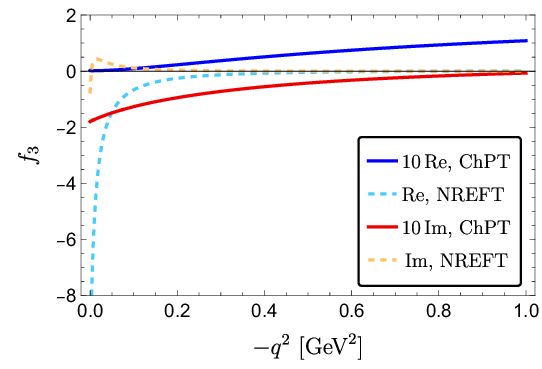}

	\caption{Comparison between the perturbative ChPT results (solid lines) and NREFT results (dashed lines). The latter are taken from Ref.~\cite{Meissner:2026zos}.
		The $f_{1}(q^2),f_{2}(q^2),f_{3}(q^2)$ in the range $-1\, \mbox{GeV}^2<q^2
		<0$ are plotted.}
	\label{fig:ChPTvsNREFT}
\end{figure}

	Another important point concerns the behavior of the form factor at small $q^2$.
	As already mentioned in the introduction, the calculations carried out in the framework
	of NREFT lead to a large curvature near origin that is a non-perturbative effect
	caused by the proximity of the resonance to the real axis.\footnote{This effect is
	reminiscent of the unnaturally large scattering length in a system that features very
		shallow bound state. We address this issue in more detail in the appendix~\ref{app:q2}.} 
		It is clear that our perturbative calculations with the vertices
	emerging from the leading-order Lagrangian will fail to reproduce this curvature.
	The Fig.~\ref{fig:ChPTvsNREFT}, where the NREFT results from
	Ref.~\cite{Meissner:2026zos} are compared to our calculations, clearly demonstrates
	this failure. It is namely seen that ChPT calculations do not capture the fine structure
	of all three form factors in the interval $-0.1\, \mbox{GeV}^2<q^2< 0$.

\begin{figure}[t]
		\centering

		\myfiggg{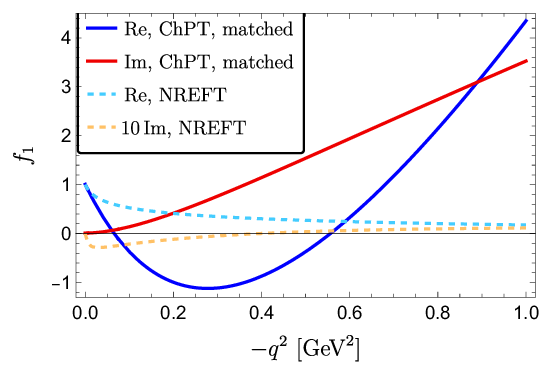}\hfill
		\myfiggg{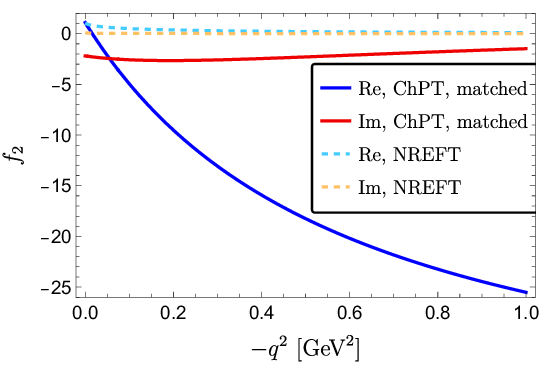}
		
		\vspace{5mm}
		
		\myfiggg{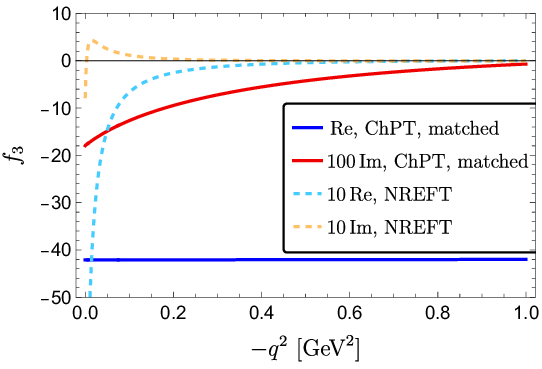}

		\caption{Comparison between the perturbative ChPT results for LECs matched to the NREFT results (solid lines) and NREFT results (dashed lines). }
		\label{fig:LECsMatched}
	\end{figure}

	Furthermore, one knows that the contribution of a resonance in a perturbative framework
	can be mimicked by unnaturally large LECs. We have the LECs $f_V,d_x,h_V$ at our
	disposal, and we may choose these to reproduce the behavior of the form factors
	near $q^2=0$. Namely, we can match the values of $D_x$ and $h_V$ to the {\em real} parts\footnote{One could equally well have matched both real and imaginary parts.
		However, let us note that we perform this matching for illustrative purpose only.
		Matching the imaginary part as well will make little difference in this case.}
	of charge radius and the magnetic moment, calculated in NREFT. As expected, these turn to be large
	\eq
	D_x=-3.41, \qquad h_V= 10.53\, .
	\en 
	The results of our calculations are shown in Fig.~\ref{fig:LECsMatched}.
	Note that the real part of the quadrupole momentum that is given by $f_3(q^2)$ is
	{\em predicted} to be large and negative now (we remind the reader that
	the quadrupole momentum is not part
	of the matching). Despite this, the results should be taken with a grain of salt, since now the
	convergence is apparently very bad beyond the small interval $q^2\approx 0$. To
	summarize, we observe a clear signature coming from the narrow resonance in the
	form factors, which cannot be reasonably reproduced in perturbative calculations at
	one loop. It remains to be seen, whether this signature can be independently
	verified by future lattice calculations. 

	\section{Conclusions}
	\label{sec:concl}
	In this paper we have calculated the electromagnetic form factors of the
	$\rho$-meson in Chiral Perturbation Theory up to one-loop. To this end,
	the method of infrared regularization (IR) was applied. In particular, the
	Feynman amplitudes were reduced to 13 independent scalar integrals, and
	the power-counting violating polynomial terms obtained by using IR were
	subtracted from all these integrals. It has been shown that this procedure
	is consistent, i.e., the subtracted amplitudes obey both chiral Ward identities
	as well as the Ward identities related to the conservation of the electromagnetic
	current  (up to the chiral order one is working). Analytic continuation to the second Riemann sheet is discussed in detail. Final results were compared with
	the recent calculations carried out in NREFT~\cite{Meissner:2026zos}.
	
	In brief, the findings of the present paper could be summarized as follows:

	\begin{itemize}
		\item
		The use of IR renders the application of the Chiral Perturbation Theory to the calculation
		of the $\rho$-meson form factors consistent. This statement remains true even   though one
		here performs subtractions at different momentum scales: at momenta of   the order of
		$m_\rho^2$ as well as soft momenta of  the order of $q^2=O(M^2)$. It is argued that
		the effective Lagrangian which includes terms with direct couplings of the photon   to the $\rho$-meson
		possesses sufficient freedom to accommodate all these subtractions.
		
		\item
		As expected, the convergence of the chiral expansion in the theory with IR is
		improved substantially.

		\item
		The salient feature of the form factors, calculated within NREFT~\cite{Meissner:2026zos},
		is the rapid variation near origin. Namely, both the charge radius
		and the quadrupole moment of the $\rho$-meson, which measure
		charge distribution inside this resonance, turn out to be unnaturally large. The
		magnetic moment in NREFT is also smaller (of  the order of 1) than in most of the
		phenomenological approaches. These predictions, which can be traced back to the
		non-perturbative effects in the vicinity of a narrow resonance, are rather robust
		in the assumption of the NREFT effective couplings of natural size, and can be verified
		independently, e.g., in lattice QCD calculations.
		
		\item
		The predictions from ChPT strongly differ from the above picture. Namely, ChPT
		fails to reproduce the structure of the form factors at small $q^2$. Formally,
		one could match these two theories at $q^2=0$ at the expense of unnaturally large LECs.
		In any case, as seen from comparison, ChPT does not converge for the values
		$q^2<-0.1\,\mbox{GeV}^2$ since  it still fails to reproduce rapid variations in the form factors.
		
		\item
		The above-mentioned salient feature of the form factors is universal (i.e.,  it does not
		depend  on the resonance considered) and stems from the vicinity of the resonance pole
		to the real axis. Moreover, this feature is not captured by perturbative calculations
		and might be important from the point of view of phenomenology.
		Therefore, an independent verification of this prediction on the lattice would be very
		timely.

	\end{itemize}

	{\em Acknowledgments:}  
	The authors thank Christoph Hanhart and
        Ajay Sakhthivasan for interesting discussions.
	The work of A.R. was  funded in part by Deutsche Forschungsgemeinschaft
	(DFG, German Research Foundation)  -- Project number RU 1205/2-1,
	by the Ministry of Culture and Science of North Rhine-Westphalia through the
	NRW-FAIR project and by the Chinese Academy of Sciences (CAS) President's
        International Fellowship Initiative (PIFI) grant no. 2024VMB0001.
        The work of U.G.M. was supported in part 
	by Deutsche Forschungsgemeinschaft (DFG) under Germany's Excellence Strategy -- EXC 3107 -- Project-ID~533766364 and  by the CAS President's International Fellowship Initiative (PIFI) under Grant No. 2025PD0022.
	The work by J.G. was supported in part by the MKW NRW under the funding code NW21-024-A, and by the European Research Council (ERC) under the European Union’s Horizon 2020 research and innovation programm (grant agreement No. 885150).

	\newpage

	\appendix
	
	\section{Explicit expressions}\label{app:individual}
	In this appendix, we list the full expressions for the quantities we have calculated
	in this paper.
	We begin with the expressions for the form factors coming from one-loop diagrams (3)-(22) in Fig.~\ref{fig:Diagrams}. All of the formulas presented below are obtained in $D$ dimensions {\em before} subtracting the  infrared regular parts. 
	
	The contributions of individual one-loop diagrams to $f_1(q^2)$ read:
	\eq
	f_1^{(3)}(q^2) &=& -\frac{e g^2 m_\rho^2}{8\pi^2(D-2)(D-1)(q^2-4m_\rho^2)^2(q^2-m_\rho^2)}\Bigl\{ \nonumber\\
	&&\Bigl[m_\rho^2\bigl(16(D-2)M^2-Dq^2\bigr)     -4(D-2)M^2q^2+4m_\rho^4\Bigr]B_0(m_\rho^2,M^2,M^2) \nonumber\\	
	&&+\Bigl[-4m_\rho^2\bigl(4(D-2)M^2+q^2\bigr)+4(D-2)M^2q^2\nonumber\\
	&&+4(D-1)m_\rho^4+q^4\Bigr]B_0(q^2,M^2,M^2)\nonumber\\
	&&+2(D-1)(q^2-2m_\rho^2)(-4M^2m_\rho^2\nonumber\\
	&&+M^2q^2+m_\rho^4)C_0(m_\rho^2,m_\rho^2,q^2,M^2,M^2,M^2)\Bigr\},\\ [10pt]
	f_1^{(4)}(q^2) &=& \frac{e g^2}{128\pi^2(D-2)(D-1)m_\rho^2(q^2-4m_\rho^2)^2(m_\rho^2-q^2)}\Bigl\{ \nonumber\\
	&&-2(D-2)(q^2-4m_\rho^2)^2\bigl[(12D-17)m_\rho^2+2q^2\bigr]A_0(m_\rho^2)\nonumber\\
	&&+m_\rho^4\Bigl[-4(108D^2-317D+206)q^2m_\rho^2+16(44D^2-137D+99)m_\rho^4 \nonumber\\
	&&+(60D^2-167D+98)q^4\Bigr]B_0(m_\rho^2,m_\rho^2,m_\rho^2) +2\Bigl[(8D^2-38D+43)q^6m_\rho^2\nonumber\\
	&&-2(46D^2-176D+165)q^4m_\rho^4+4(80D^2-291D+259)q^2m_\rho^6\nonumber\\
	&&-8(36D^2-137D+129)m_\rho^8+(D-2)q^8\Bigr]B_0(q^2,m_\rho^2,m_\rho^2)\nonumber\\
	&&-4(D-1)m_\rho^4\Bigl[(53-30D)q^4m_\rho^2 +8(13D-24)q^2m_\rho^4\nonumber\\
	&&+(276-144D)m_\rho^6+(3D-5)q^6\Bigr]C_0(m_\rho^2,m_\rho^2,q^2,m_\rho^2,m_\rho^2,m_\rho^2)\Bigr\},\\ [10pt]
	f_1^{(5)}(q^2) &=& 0,\\ [10pt]
	f_1^{(6)}(q^2) &=& \frac{(D-1)^2 e g^2 m_\rho^2}{8\pi^2 D(q^2-m_\rho^2)^2}A_0(m_\rho^2),\\ [10pt]
	f_1^{(7)}(q^2) &=& \frac{e g^2}{64\pi^2(D-1)D m_\rho^2(q^2-m_\rho^2)^2}\Bigl\{\nonumber\\
	&&-2\bigl[4(D-2)Dq^2m_\rho^2+4(D-1)m_\rho^4+Dq^4\bigr]A_0(m_\rho^2)\nonumber\\
	&&+D(q^2-4m_\rho^2)\bigl[4(2D-3)q^2m_\rho^2+4(D-1)m_\rho^4+q^4\bigr]B_0(q^2,m_\rho^2,m_\rho^2)\Bigr\}, \\ [10pt]
	f_1^{(8+9)}(q^2) &=& -\frac{3 e g^2}{128\pi^2(D-1)(q^2-m_\rho^2)}
	\Bigl\{\nonumber\\
	&&2A_0(m_\rho^2)+3(5-4D)m_\rho^2 B_0(m_\rho^2,m_\rho^2,m_\rho^2)\Bigr\},\\ [10pt]
	f_1^{(10)}(q^2) &=& \frac{e g^2 m_\rho^2}{16\pi^2(D-1)(q^2-m_\rho^2)^2}
	\Bigl\{-2A_0(M^2)+(q^2-4M^2)B_0(q^2,M^2,M^2)\Bigr\},\\ [10pt]
	f_1^{(11)}(q^2) &=& -\frac{e m_\rho^2}{16\pi^2F^2(q^2-m_\rho^2)}A_0(M^2),\\ [10pt]
	f_1^{(12)}(q^2) &=&\frac{(D-3)(D-2)e g_{\omega\rho\pi}^2 m_\rho^2}
	{64\pi^2(D-1)F^2(q^2-m_\rho^2)^2}\Bigl\{\nonumber\\
	&&(M^2-m_\omega^2-q^2)A_0(m_\omega^2)-(M^2-m_\omega^2+q^2)A_0(M^2) \nonumber\\
	&&+\Bigl[(M^2-q^2)^2-2(M^2+q^2)m_\omega^2+m_\omega^4\Bigr]B_0(q^2,M^2,m_\omega^2)\Bigr\},\\[10pt]
	f_1^{(13)}(q^2) &=&-\frac{(D-3)e g_{\omega\rho\pi}^2}{128(D-1)F^2\pi^2(q^2-4m_\rho^2)^2(q^2-m_\rho^2)}
	\Bigl\{\nonumber\\
	&&-2(q^2-4m_\rho^2)\Bigl[-2m_\rho^4+\bigl((6-4D)M^2+q^2+(4D-6)m_\omega^2\bigr)m_\rho^2\nonumber\\
	&&+(D-2)q^2(M^2-m_\omega^2)\Bigr]A_0(m_\omega^2)+2(q^2-4m_\rho^2)\Bigl[2(2D-5)m_\rho^4\nonumber\\
	&&+\bigl((6-4D)M^2-(D-3)q^2+(4D-6)m_\omega^2\bigr)m_\rho^2\nonumber\\ &&+(D-2)q^2(M^2-m_\omega^2)\Bigr]A_0(M^2) \nonumber\\
	&&-\frac{1}{D-2}\Bigl[-16(D^2-4D+3)m_\rho^8+4\bigl(8(D-3)^2M^2+(D^2-D-8)q^2 \nonumber\\
	&&+8(D^2-4D+3)m_\omega^2\bigr)m_\rho^6-4\Bigl(4(D^2-4D+3)M^4 \nonumber\\
	&&+2(D^2-9D+18)q^2M^2+(2D-7)q^4+4(D^2-4D+3)m_\omega^4\nonumber\\
	&&-2(D-1)\bigl(4(D-3)M^2+(14-5D)q^2\bigr)m_\omega^2\Bigr)m_\rho^4\nonumber\\
	&&+q^2\Bigl(4(D^2-5D+4)m_\omega^4-8(D-1)\bigl((D-4)M^2-(D-3)q^2\bigr)m_\omega^2 \nonumber\\
	&&+(D-4)(4(D-1)M^4-4q^2M^2+q^4)\Bigr)m_\rho^2\Bigr]B_0(q^2,M^2,M^2) \nonumber\\
	&&-\frac{2}{D-2}\Bigl[-8(D^2-3D+1)m_\rho^8-2\bigl(8(D-2)^2M^2+(-3D^2+10D-4)q^2 \nonumber\\
	&&+8(D-2)^2m_\omega^2\bigr)m_\rho^6+\Bigl(8(3D^2-11D+9)M^4+4(D^2-5D+7)q^2M^2 \nonumber\\
	&&-(D^2-4D+2)q^4+8(3D^2-11D+9)m_\omega^4-4\bigl(4(3D^2-11D+9)M^2\nonumber\\
	&&+(-5D^2+19D-17)q^2\bigr)m_\omega^2\Bigr)m_\rho^4 -2q^2\Bigl((5D^2-20D+18)m_\omega^4\nonumber\\
	&&+\bigl((2D^2-8D+7)q^2 -2(5D^2-20D+18)M^2\bigr)m_\omega^2\nonumber\\
	&&+(D-2)^2q^4(M^2-m_\omega^2)^2\nonumber\\
	&&+M^2\bigl((5D^2-20D+18)M^2+q^2\bigr)\Bigr)m_\rho^2\Bigr]B_0(m_\rho^2,M^2,m_\omega^2) \nonumber\\
	&&+\frac{2(D-1)m_\rho^2}{D-2}(2M^2-q^2+2m_\rho^2-2m_\omega^2)\Bigl[-4(D-3)m_\rho^6 +\bigl(8(D-3)M^2\nonumber\\
	&&+(D-4)q^2+8(D-3)m_\omega^2\bigr)m_\rho^4\nonumber\\
	&&-2\Bigl(2(D-3)M^4+(D-4)q^2M^2+2(D-3)m_\omega^4\nonumber\\
	&& +\bigl((5D-14)q^2-4(D-3)M^2\bigr)m_\omega^2\Bigr)m_\rho^2 +q^2\Bigl((D-4)M^4+(D-4)m_\omega^4\nonumber\\
	&&+\bigl(2(D-3)q^2-2(D-4)M^2\bigr)m_\omega^2\Bigr)\Bigr]C_0(m_\rho^2,m_\rho^2,q^2,M^2,m_\omega^2,M^2)\Bigr\},\\[10pt]
	f_1^{(14+15)}(q^2) &=&0,\\  [10pt]
	f_1^{(16+17)}(q^2) &=&-\frac{(D-3)(D-2)e g_{\omega\rho\pi}^2}{64\pi^2(D-1)F^2(q^2-m_\rho^2)}\Bigl\{\nonumber\\
	&&(M^2-m_\rho^2-m_\omega^2)A_0(m_\omega^2) -(M^2+m_\rho^2-m_\omega^2)A_0(M^2)\nonumber\\
	&&+\Bigl[-2m_\rho^2(M^2+m_\omega^2)+(M^2-m_\omega^2)^2+m_\rho^4\Bigr] B_0(m_\rho^2,M^2,m_\omega^2)\Bigr\},\\[10pt]
	f_1^{(18)}(q^2) &=&0,\\[10pt]
	f_1^{(19+20)}(q^2) &=&
	\frac{(D-3)e g_{\omega\rho\pi}^2m_\rho^2}
	{32(D-2)(D-1)F^2\pi^2q^2(q^2-4m_\rho^2)^2(q^2-m_\rho^2)}\Bigl\{ \nonumber\\
	&&(D-2)q^2(q^2-4m_\rho^2)(M^2+q^2-2m_\rho^2-m_\omega^2)A_0(m_\omega^2) \nonumber\\
	&&-(D-2)q^2(q^2-4m_\rho^2)(M^2+q^2-2m_\rho^2-m_\omega^2)A_0(M^2) \nonumber\\
	&&+q^2\Bigl[4(D-1)(2M^2-q^2-2m_\omega^2)m_\rho^4\nonumber\\
	&&-2\Bigl(2M^4+(7D-11)q^2M^2 +(5-3D)q^4\nonumber\\
	&&+2m_\omega^4+\bigl((D-5)q^2-4M^2\bigr)m_\omega^2\Bigr)m_\rho^2 +q^2\Bigl(2q^4-6M^2q^2+Dm_\omega^4\nonumber\\
	&&-2(DM^2+q^2)m_\omega^2+D(M^4+4q^2M^2-q^4)\Bigr)\Bigr] B_0(q^2,M^2,m_\omega^2) \nonumber\\
	&&+\Bigl[-4(D-2)M^2q^6+8(D-2)m_\rho^6q^2+m_\rho^2\Bigl(-4(D-1)M^4 \nonumber\\
	&&+2(11D-23)q^2M^2+(D-2)q^4-4(D-1)m_\omega^4\nonumber\\
	&&+2(D-1)(4M^2+q^2)m_\omega^2\Bigr)q^2 +2m_\rho^4\Bigl(4(D-1)M^4-16(D-2)q^2M^2 \nonumber\\
	&&-8(D-1)m_\omega^2M^2+(7-4D)q^4+4(D-1)m_\omega^4\Bigr)\Bigr]B_0(m_\rho^2,M^2,M^2) \nonumber\\
	&&-2\Bigl[4(D-2)q^2m_\rho^6+\Bigl(4(D-1)M^4-4(3D-5)q^2M^2\nonumber\\
	&&+q^4+4(D-1)m_\omega^4 -4\bigl(2(D-1)M^2+(D-3)q^2\bigr)m_\omega^2\Bigr)m_\rho^4 \nonumber\\
	&&-2q^2\Bigl(2M^4+(3-2D)q^2M^2+2m_\omega^4+(q^2-4M^2)m_\omega^2\Bigr)m_\rho^2\nonumber\\
	&& +q^4(M^2-m_\omega^2)^2\Bigr]B_0(m_\rho^2,M^2,m_\omega^2) \nonumber\\
	&&+4(D-1)\Bigl[M^2(M^2-m_\omega^2)q^6+m_\rho^6(-2M^2+q^2+2m_\omega^2)q^2 \nonumber\\
	&&+m_\rho^2\Bigl(-m_\omega^6+(3M^2+q^2)m_\omega^4+(4M^2q^2-3M^4)m_\omega^2\nonumber\\
	&&+M^2(M^4-5q^2M^2+q^4)\Bigr)q^2 \nonumber\\
	&&-m_\rho^4\Bigl(2M^6-7q^2M^4+2q^4M^2-2m_\omega^6+(6M^2+q^2)m_\omega^4 \nonumber\\
	&&+(-6M^4+6q^2M^2+2q^4)m_\omega^2\Bigr)\Bigr] C_0(m_\rho^2,m_\rho^2,q^2,M^2,M^2,m_\omega^2)\Bigr\},\\ [10pt]
	f_1^{(21+22)}(q^2) &=& 0.
	\en
	
	\bigskip
	
	The contributions of the one-loop diagrams to $f_2(q^2)$ read:
	
	\eq
	f_2^{(3)}(q^2) &=& \frac{e g^2 m_\rho^2}{8\pi^2(D-2)(D-1)(q^2-4m_\rho^2)^2(q^2-m_\rho^2)}\Bigl\{\nonumber\\ &&\Bigl[(D-2)(16M^2+q^2)m_\rho^2-4(D-2)M^2q^2\nonumber\\
	&&+(12-8D)m_\rho^4\Bigr]B_0(m_\rho^2,M^2,M^2)\nonumber\\
	&&+\Bigl[-2m_\rho^2\bigl(8(D-2)M^2+(5-3D)q^2\bigr)-4(D-1)m_\rho^4\nonumber\\
	&&-(D-2)q^2(q^2-4M^2)\Bigr]B_0(q^2,M^2,M^2)\nonumber\\
	&&+4(D-1)m_\rho^2(m_\rho^4-4M^2m_\rho^2+M^2q^2)C_0(m_\rho^2,m_\rho^2,q^2,M^2,M^2,M^2)\Bigr\},\\ [10pt]
	f_2^{(4)}(q^2) &=&\frac{e g^2}{128\pi^2(D-2)(D-1)m_\rho^2(q^2-4m_\rho^2)^2(m_\rho^2-q^2)}\Bigl\{ \nonumber\\
	&&-2(D-2)(q^2-4m_\rho^2)^2\bigl[(24D-31)m_\rho^2+q^2\bigr]A_0(m_\rho^2)\nonumber\\
	&&+m_\rho^4\Bigl[-4(100D^2-265D+132)q^2m_\rho^2+16(28D^2-59D+7)m_\rho^4\nonumber\\
	&&+(76D^2-219D+134)q^4\Bigr]B_0(m_\rho^2,m_\rho^2,m_\rho^2) \nonumber\\
	&&+\Bigl[4(5D^2-19D+18)q^6m_\rho^2\nonumber\\
	&&-4(50D^2-170D+139)q^4m_\rho^4+16(28D^2-91D+69)q^2m_\rho^6 \nonumber\\
	&&+16(4D^2-15D+15)m_\rho^8+(D-2)q^8\Bigr]B_0(q^2,m_\rho^2,m_\rho^2)\nonumber\\
	&&-8(D-1)m_\rho^4\Bigl[(33-17D)q^4m_\rho^2+(68D-131)q^2m_\rho^4\nonumber\\
	&&-6(20D-39)m_\rho^6+2(D-2)q^6\Bigr]C_0(m_\rho^2,m_\rho^2,q^2,m_\rho^2,m_\rho^2,m_\rho^2)\Bigr\},\\[10pt]
	f_2^{(5)}(q^2) &=&\frac{3e g^2}{128\pi^2(D-1)m_\rho^2(m_\rho^2-q^2)}
	\Bigl\{-2(q^2-2m_\rho^2)A_0(m_\rho^2) \nonumber\\
	&&+(q^2-4m_\rho^2)\bigl[(4D-6)m_\rho^2+q^2\bigr]B_0(q^2,m_\rho^2,m_\rho^2)\Bigr\},\\[10pt]
	f_2^{(6)}(q^2) &=&\frac{(D-1)^2e g^2m_\rho^2}{4\pi^2D(q^2-m_\rho^2)^2}A_0(m_\rho^2),\\[10pt]
	f_2^{(7)}(q^2) &=&\frac{e g^2}{32\pi^2(D-1)Dm_\rho^2(q^2-m_\rho^2)^2}\Bigl\{\nonumber\\
	&&-2\bigl[4(D-2)Dq^2m_\rho^2+4(D-1)m_\rho^4+Dq^4\bigr]A_0(m_\rho^2) \nonumber\\
	&&+D(q^2-4m_\rho^2)\bigl[4(2D-3)q^2m_\rho^2\nonumber\\
	&&+4(D-1)m_\rho^4+q^4\bigr]B_0(q^2,m_\rho^2,m_\rho^2)\Bigr\},\\ [10pt]
	f_2^{(8+9)}(q^2) &=&-\frac{3eg^2}{128\pi^2(D-1)(q^2-m_\rho^2)}
	\Bigl\{\nonumber\\
	&&2A_0(m_\rho^2)+3(5-4D)m_\rho^2B_0(m_\rho^2,m_\rho^2,m_\rho^2)\Bigr\},\\[10pt]
	f_2^{(10)}(q^2) &=&\frac{eg^2m_\rho^2}{8\pi^2(D-1)(q^2-m_\rho^2)^2}\Bigl\{-2A_0(M^2)
	+(q^2-4M^2)B_0(q^2,M^2,M^2)\Bigr\},\\[10pt]
	f_2^{(11)}(q^2) &=&-\frac{em_\rho^2}{8\pi^2F^2(q^2-m_\rho^2)}A_0(M^2),\\[10pt]
	f_2^{(12)}(q^2) &=&\frac{(D-3)(D-2)eg_{\omega\rho\pi}^2m_\rho^2}{32\pi^2(D-1)F^2(q^2-m_\rho^2)^2}
	\Bigl\{\nonumber\\
	&&(M^2-m_\omega^2-q^2)A_0(m_\omega^2)-(M^2-m_\omega^2+q^2)A_0(M^2) \nonumber\\
	&&+\Bigl[(M^2-q^2)^2-2(M^2+q^2)m_\omega^2+m_\omega^4\Bigr]B_0(q^2,M^2,m_\omega^2)\Bigr\},\\ [10pt]
	f_2^{(13)}(q^2) &=&-\frac{(D-3)eg_{\omega\rho\pi}^2m_\rho^2}{32\pi^2(D-1)F^2(q^2-4m_\rho^2)^2(q^2-m_\rho^2)}\Bigl\{\nonumber\\
	&&(q^2-4m_\rho^2)(-M^2+m_\rho^2+m_\omega^2)A_0(m_\omega^2)\nonumber\\ &&-(q^2-4m_\rho^2)(-M^2+m_\rho^2+m_\omega^2)A_0(M^2) \nonumber\\
	&&-\frac{1}{D-2}
	\Bigl[m_\rho^4\bigl(8(2D-3)M^2+(D-2)q^2-8m_\omega^2\bigr)+(D-2)q^2(M^2-m_\omega^2)^2\nonumber\\
	&&-2m_\rho^2\Bigl((4D-6)M^4+m_\omega^2\bigl((12-8D)M^2+(D-2)q^2\bigr) \nonumber\\
	&&+(D-2)M^2q^2+(4D-6)m_\omega^4\Bigr)+(12-8D)m_\rho^6\Bigr]B_0(m_\rho^2,M^2,m_\omega^2) \nonumber\\
	&&+\frac{m_\rho^2}{D-2}\Bigl[2m_\rho^2\bigl(8(D-2)M^2+(5-3D)q^2\bigr)+2(D-1)(4M^2-q^2)m_\omega^2\nonumber\\
	&&-4DM^4 -2DM^2q^2+4(D-1)m_\rho^4-4(D-1)m_\omega^4+Dq^4\nonumber\\
	&&+4M^4+6M^2q^2-2q^4\Bigr]B_0(q^2,M^2,M^2) \nonumber\\
	&&-\frac{4(D-1)m_\rho^2}{D-2}(-M^2+m_\rho^2+m_\omega^2)\Bigl[(q^2-2M^2)m_\omega^2
	-2m_\rho^2(M^2+m_\omega^2) \nonumber\\
	&&+M^4+m_\rho^4+m_\omega^4\Bigr]C_0(m_\rho^2,m_\rho^2,q^2,M^2,m_\omega^2,M^2)\Bigr\}, \\[10pt]
	f_2^{(14+15)}(q^2) &=&0,\\  [10pt]
	f_2^{(16+17)}(q^2) &=&-\frac{(D-3)(D-2)eg_{\omega\rho\pi}^2}{64\pi^2(D-1)F^2(q^2-m_\rho^2)}
	\Bigl\{\nonumber\\
	&&(M^2-m_\rho^2-m_\omega^2)A_0(m_\omega^2)-(M^2+m_\rho^2-m_\omega^2)A_0(M^2) \nonumber\\
	&&+\Bigl[-2m_\rho^2(M^2+m_\omega^2)+(M^2-m_\omega^2)^2+m_\rho^4\Bigr]B_0(m_\rho^2,M^2,m_\omega^2)\Bigr\},\\[10pt]
	f_2^{(18)}(q^2) &=&-\frac{(D-3)(D-2)eg_{\omega\rho\pi}^2m_\rho^2}
	{64\pi^2(D-1)F^2q^2(q^2-m_\rho^2)}\Bigl\{\nonumber\\
	&&(M^2-m_\omega^2-q^2)A_0(m_\omega^2)-(M^2-m_\omega^2+q^2)A_0(M^2) \nonumber\\
	&&+\Bigl[(M^2-q^2)^2-2(M^2+q^2)m_\omega^2+m_\omega^4\Bigr]B_0(q^2,M^2,m_\omega^2)\Bigr\},
	\\[10pt]
	f_2^{(19+20)}(q^2)&=&-\frac{(D-3)eq^2g_{\omega\rho\pi}^2}{64(D-2)(D-1)F^2\pi^2(q^2-m_\rho^2)\bigl(q^4-4q^2m_\rho^2\bigr)^3}\Bigl\{ \nonumber\\
	&&(q^2-4m_\rho^2)\bigl(q^4-4q^2m_\rho^2\bigr)\Bigl[-(D-2)^2(M^2-m_\omega^2)q^4 \nonumber\\
	&&+(3D^2-10D+8)m_\rho^2(M^2-m_\omega^2)q^2+4(D-2)m_\rho^4\bigl((D-2)M^2 \nonumber\\
	&&+q^2-(D-2)m_\omega^2\bigr)\Bigr]A_0(m_\omega^2)\nonumber\\
	&&+(q^2-4m_\rho^2)\bigl(q^4-4q^2m_\rho^2\bigr)\Bigl\{(2-D)\Bigl[-(D-2)(M^2-m_\omega^2)q^4 \nonumber\\
	&&+(D-2)m_\rho^2(3M^2+2q^2-3m_\omega^2)q^2 \nonumber\\
	&&+2m_\rho^4\bigl(2(D-1)M^2+(9-4D)q^2-2(D-1)m_\omega^2\bigr)\Bigr]\nonumber\\
	&&-2(D-2)m_\rho^2\Bigl[(M^2-m_\omega^2)q^2+m_\rho^2(-2M^2+q^2+2m_\omega^2)\Bigr]\Bigr\}A_0(M^2) \nonumber\\
	&&+q^2m_\rho^2(q^2-4m_\rho^2)\Bigl\{4(D-2)M^2q^6-(D-2)m_\rho^2\Bigl[4(D-1)M^4 \nonumber\\
	&&-4(D-13)q^2M^2+q^4+4(D-1)m_\omega^4-4(D-1)(2M^2+q^2)m_\omega^2\Bigr]q^2 \nonumber\\
	&&-8m_\rho^6\Bigl[2(D^2-4D+3)M^2+(-2D^2+9D-11)q^2\nonumber\\
	&&-2(D^2-4D+3)m_\omega^2\Bigr] \nonumber\\
	&&+2m_\rho^4\Bigl[8(D^2-4D+3)M^4-2(3D^2-36D+65)q^2M^2\nonumber\\
	&&+(-2D^2+9D-13)q^4 +8(D^2-4D+3)m_\omega^4\nonumber\\
	&&-2(D^2-4D+3)(8M^2+5q^2)m_\omega^2
	\Bigr]\Bigr\}B_0(m_\rho^2,M^2,M^2) \nonumber\\
	&&+m_\rho^2\bigl(q^4-4q^2m_\rho^2\bigr)\Bigl\{(D-2)^2\Bigl[m_\omega^4
	-2(M^2+q^2)m_\omega^2+(M^2-q^2)^2\Bigr]q^4 \nonumber\\
	&&+2m_\rho^2\Bigl[(4D-6)M^4+(3D^2-10D+11)q^2M^2\nonumber\\
	&&+(-3D^2+12D-11)q^4 \nonumber\\
	&&+(4D-6)m_\omega^4+\bigl((12-8D)M^2+(5D^2-22D+21)q^2\bigr)m_\omega^2
	\Bigr]q^2 \nonumber\\
	&&-4(D-1)m_\rho^4(2M^2+q^2-2m_\omega^2)\bigl[2(D-2)M^2\nonumber\\
	&&+(5-2D)q^2-2(D-2)m_\omega^2\bigr] \nonumber\\
	&&+4m_\rho^2\Bigl[4(D-1)q^2m_\rho^4+4\Bigl((D-2)M^4+(5-3D)q^2M^2-q^4 \nonumber\\
	&&+(D-2)m_\omega^4+\bigl(-2(D-2)M^2-(D-3)q^2\bigr)m_\omega^2\Bigr)m_\rho^2 \nonumber\\
	&&+q^2\Bigl(M^4+2(2D-3)q^2M^2+q^4+m_\omega^4\nonumber\\
	&&-2(M^2+q^2)m_\omega^2\Bigr)
	\Bigr]\Bigr\}B_0(q^2,M^2,m_\omega^2) \nonumber\\
	&&+q^2(q^2-4m_\rho^2)\Bigl\{8(D-1)\bigl[2(D-3)M^2+q^2-2(D-3)m_\omega^2\bigr]m_\rho^8 \nonumber\\
	&&-2\Bigl[8(D^2-4D+3)M^4-2(D^2-1)q^2M^2+(-2D^2+11D-11)q^4 \nonumber\\
	&&+8(D^2-4D+3)m_\omega^4-2(D-1)\bigl(8(D-3)M^2\nonumber\\
	&&+(7D-17)q^2\bigr)m_\omega^2\Bigr]m_\rho^6 \nonumber\\
	&&-q^2\Bigl[4q^4+30M^2q^2+4(D-1)Dm_\omega^4+\bigl((14q^2-8M^2)D^2 \nonumber\\
	&&+8(M^2-7q^2)D+50q^2\bigr)m_\omega^2-4D(M^4+8q^2M^2+q^4) \nonumber\\
	&&+D^2(4M^4+10q^2M^2+q^4)\Bigr]m_\rho^4+2(D-2)q^4\Bigl[(3D-5)M^4 \nonumber\\
	&&+(D-2)q^2M^2+(3D-5)m_\omega^4\nonumber\\
	&&+\bigl((10-6D)M^2+(D-2)q^2\bigr)m_\omega^2\Bigr]m_\rho^2 \nonumber\\
	&&-(D-2)^2q^6(M^2-m_\omega^2)^2\Bigr\}B_0(m_\rho^2,M^2,m_\omega^2) \nonumber\\
	&&+2(D-1)q^2m_\rho^4(q^2-4m_\rho^2)\Bigl\{4(M^2+q^2-2m_\rho^2-m_\omega^2)\Bigl[M^2q^4+m_\rho^4q^2 \nonumber\\
	&&+m_\rho^2\bigl(M^4-3q^2M^2+m_\omega^4-(2M^2+q^2)m_\omega^2\bigr)\Bigr] \nonumber\\
	&&+(2M^2+q^2-2m_\omega^2)\Bigl[2\bigl(2(D-2)M^2+(5-2D)q^2\nonumber\\
	&&-2(D-2)m_\omega^2\bigr)m_\rho^4 \nonumber\\
	&&+\Bigl((10-4D)M^4+3(D-4)q^2M^2+(D-2)q^4+(10-4D)m_\omega^4 \nonumber\\
	&&+\bigl(4(2D-5)M^2+(5D-12)q^2\bigr)m_\omega^2\Bigr)m_\rho^2 \nonumber\\
	&&+q^2\Bigl((D-2)M^4-(D-4)q^2M^2+(D-2)m_\omega^4\nonumber\\
	&&-(D-2)(2M^2+q^2)m_\omega^2
	\Bigr)\Bigr]\Bigr\} C_0(m_\rho^2,m_\rho^2,q^2,M^2,M^2,m_\omega^2)\Bigr\},\\ [10pt]
	f_2^{(21+22)}(q^2) &=&0. 
	\en
	
	\bigskip
	
	The contributions of the one-loop diagrams to $f_3(q^2)$ read:

	\eq
	f_3^{(3)}(q^2) &=&-\frac{eg^2m_\rho^2}{8\pi^2(D-2)(D-1)(q^2-4m_\rho^2)^3(q^4-q^2m_\rho^2)}\Bigl\{
	\nonumber\\
	&&-2(D-2)^2(q^2-4m_\rho^2)^2A_0(M^2)\nonumber\\
	&&-2\Bigl[4(D^2-3D+1)q^2m_\rho^4-8(D^2-4D+3)m_\rho^6\nonumber\\
	&&+m_\rho^2(Dq^4-16(D-2)M^2q^2)+4(D-2)M^2q^4\Bigr]B_0(m_\rho^2,M^2,M^2) \nonumber\\
	&&+\Bigl[4m_\rho^4\bigl((D^2-3D+6)q^2+16(D-2)M^2\bigr)\nonumber\\
	&&+8q^2m_\rho^2\bigl((D-3)q^2-8(D-2)M^2\bigr) \nonumber\\
	&&+12(D-2)M^2q^4-16(D-1)m_\rho^6-(D-4)q^6\Bigr]B_0(q^2,M^2,M^2) \nonumber\\
	&&+2(D-1)\Bigl[-2m_\rho^6(Dq^2+16M^2)+m_\rho^4(Dq^4+24M^2q^2) \nonumber\\
	&&-12M^2q^4m_\rho^2+2M^2q^6+8m_\rho^8\Bigr]C_0(m_\rho^2,m_\rho^2,q^2,M^2,M^2,M^2)\Bigr\},\\[10pt]
	f_3^{(4)}(q^2) &=&-\frac{eg^2\bigl(q^2+(14-8D)m_\rho^2\bigr)}{64\pi^2(D-2)(D-1)q^2(q^2-4m_\rho^2)^3(m_\rho^2-q^2)}\Bigl\{\nonumber\\
	&&-2(D-2)^2(q^2-4m_\rho^2)^2A_0(m_\rho^2)+2m_\rho^2\Bigl[-4(D^2-7D+9)q^2m_\rho^2 \nonumber\\
	&&+8(D^2-4D+3)m_\rho^4+(8-5D)q^4
	\Bigr]B_0(m_\rho^2,m_\rho^2,m_\rho^2) \nonumber\\
	&&+\Bigl[4(D^2-19D+38)q^2m_\rho^4+4(5D-12)q^4m_\rho^2+16(3D-7)m_\rho^6\nonumber\\
	&&-(D-4)q^6
	\Bigr]B_0(q^2,m_\rho^2,m_\rho^2) \nonumber\\
	&&+2(D-1)m_\rho^2\Bigl[(D-12)q^4m_\rho^2-2(D-12)q^2m_\rho^4\nonumber\\
	&&-24m_\rho^6+2q^6
	\Bigr]C_0(m_\rho^2,m_\rho^2,q^2,m_\rho^2,m_\rho^2,m_\rho^2)\Bigr\},\\[10pt]
	f_3^{(5)}(q^2) &=&0,\\ [10pt] 
	f_3^{(6)}(q^2) &=&0,\\ [10pt] 
	f_3^{(7)}(q^2) &=&0,\\ [10pt]
	f_3^{(8+9)}(q^2) &=&0,\\ [10pt] 
	f_3^{(10)}(q^2) &=&0,\\ [10pt]
	f_3^{(11)}(q^2) &=&0,\\ [10pt]
	f_3^{(12)}(q^2) &=&0,\\ [10pt]    
	f_3^{(13)}(q^2) &=&\frac{(D-3)eg_{\omega\rho\pi}^2}{64(D-1)F^2\pi^2q^2(q^2-4m_\rho^2)^3(q^2-m_\rho^2)}
	\Bigl\{\nonumber\\
	&&2(q^2-4m_\rho^2)\Bigl[4(D-1)m_\rho^6 \nonumber\\
	&&+\bigl(4(D-1)M^2-2q^2-4(D-1)m_\omega^2\bigr)m_\rho^4+q^2\bigl((8-6D)M^2+q^2\nonumber\\
	&&+(6D-8)m_\omega^2\bigr)m_\rho^2 +(D-2)q^4(M^2-m_\omega^2)\Bigr]A_0(m_\omega^2) \nonumber\\
	&&-2(q^2-4m_\rho^2)\Bigl[-4(D-3)m_\rho^6+2\bigl(2(D-1)M^2+(3D-7)q^2 \nonumber\\
	&&-2(D-1)m_\omega^2\bigr)m_\rho^4+q^2\bigl((8-6D)M^2-(D-3)q^2+(6D-8)m_\omega^2\bigr)m_\rho^2\nonumber\\
	&&+(D-2)q^4(M^2-m_\omega^2)
	\Bigr]A_0(M^2)\nonumber\\
	&& +\frac{m_\rho^2}{D-2}\Bigl[-32(D-1)m_\rho^8+8\bigl(8(D-3)M^2 \nonumber\\
	&&+(-3(D-5)D-8)q^2
	+8(D-1)m_\omega^2\bigr)m_\rho^6+4\Bigl(-8(D-1)M^4 \nonumber\\
	&&+12((D-5)D+8)q^2M^2+((D-5)D-8)q^4\nonumber\\
	&&+4(D-1)m_\omega^2\bigl(4M^2+(5D-16)q^2-2m_\omega^2\bigr)\Bigr)m_\rho^4 \nonumber\\
	&&+2q^2\Bigl(-4(D-1)(3D-8)M^4-4((D-9)D+19)q^2M^2+(14-3D)q^4 \nonumber\\
	&&+4(D-1)m_\omega^2\bigl(2(3D-8)M^2+(19-7D)q^2+(8-3D)m_\omega^2\bigr)\Bigr)m_\rho^2 \nonumber\\
	&&+q^4\Bigl(4(D-1)\bigl(-2(D-4)M^2+2(D-3)q^2+(D-4)m_\omega^2\bigr)m_\omega^2 \nonumber\\
	&&+(D-4)(4(D-1)M^4-4q^2M^2+q^4)\Bigr)\Bigr]B_0(q^2,M^2,M^2) \nonumber\\
	&&+\frac{2}{D-2}\Bigl[16(D-3)(D-1)m_\rho^{10}-8\bigl(D(2D-9)+6\bigr)q^2m_\rho^8 \nonumber\\
	&&-2\Bigl(8(D-3)(D-1)M^4+4\bigl(D(3D-8)+7\bigr)q^2M^2\nonumber\\
	&&+\bigl((15-4D)D-8\bigr)q^4 +8(D-3)(D-1)m_\omega^4\nonumber\\
	&&+4\bigl((5(D-4)D+17)q^2-4(D-3)(D-1)M^2\bigr)m_\omega^2
	\Bigr)m_\rho^6 \nonumber\\
	&&+q^2\Bigl(8\bigl(D(5D-17)+13\bigr)M^4+4\bigl((D-4)D+6\bigr)q^2M^2\nonumber\\
	&&-\bigl((D-4)D+2\bigr)q^4 \nonumber\\
	&&+4m_\omega^2\bigl(-4(D(5D-17)+13)M^2+(D(7D-26)+22)q^2\nonumber\\
	&&+2(D(5D-17)+13)m_\omega^2\bigr)
	\Bigr)m_\rho^4 -2q^4\Bigl((2D-5)(3D-4)M^4+q^2M^2\nonumber\\
	&&+(2D-5)(3D-4)m_\omega^4 \nonumber\\
	&&+\bigl((2(D-4)D+7)q^2-2(2D-5)(3D-4)M^2\bigr)m_\omega^2
	\Bigr)m_\rho^2\nonumber\\
	&&+(D-2)^2q^6(M^2-m_\omega^2)^2\Bigr]B_0(m_\rho^2,M^2,m_\omega^2) \nonumber\\
	&&-\frac{2(D-1)m_\rho^2}{D-2}\Bigl[-16m_\rho^{10}+4(4M^2+(14-3D)q^2+12m_\omega^2)m_\rho^8 \nonumber\\
	&&+4\Bigl(4M^4+(3D-20)q^2M^2+(2D-7)q^4-12m_\omega^4\nonumber\\
	&&+\bigl(8M^2+(13D-48)q^2\bigr)m_\omega^2\Bigr)m_\rho^6 \nonumber\\
	&&+\Bigl(-16M^6+4(3D-2)q^2M^4+2(24-7D)q^4M^2-(D-4)q^6 \nonumber\\
	&&+2m_\omega^2\Bigl(24M^4+20(D-4)q^2M^2+5(16-5D)q^4+8m_\omega^4 \nonumber\\
	&&+\bigl(2(42-13D)q^2-24M^2\bigr)m_\omega^2\Bigr)\Bigr)m_\rho^4 \nonumber\\
	&&+2q^2\Bigl((16-6D)M^6+2(D-3)q^2M^4+(D-4)q^4M^2\nonumber\\
	&&+m_\omega^2\Bigl(6(3D-8)M^4+4(14-5D)q^2M^2+(9D-26)q^4 \nonumber\\
	&&+2m_\omega^2\bigl(3(8-3D)M^2+(9D-25)q^2+(3D-8)m_\omega^2\bigr)\Bigr)
	\Bigr)m_\rho^2 \nonumber\\
	&&-q^4(-2M^2+q^2+2m_\omega^2)\Bigl[(D-4)M^4+(D-4)m_\omega^4\nonumber\\
	&&+\bigl(2(D-3)q^2-2(D-4)M^2\bigr)m_\omega^2\Bigr]\Bigr]C_0(m_\rho^2,m_\rho^2,q^2,M^2,m_\omega^2,M^2)\Bigr\}, \\[10pt]
	f_3^{(14+15)}(q^2) &=&0,\\ [10pt] 
	f_3^{(16+17)}(q^2) &=&0,\\ [10pt]
	f_3^{(18)}(q^2) &=&0,\\ [10pt]  
	f_3^{(19+20)}(q^2) &=&-\frac{(D-3)eg_{\omega\rho\pi}^2}{32(D-2)(D-1)F^2\pi^2q^4(q^2-4m_\rho^2)^3(q^2-m_\rho^2)}\Bigl\{
	\nonumber\\
	&&(2-D)q^2(q^2-4m_\rho^2)\Bigl[8(D-1)m_\rho^6 -4\bigl((2D-3)M^2+q^2\nonumber\\
	&&+(3-2D)m_\omega^2\bigr)m_\rho^4-q^2\bigl((10-7D)M^2-2q^2+(7D-10)m_\omega^2\bigr)m_\rho^2\nonumber\\
	&& -(D-2)q^4(M^2-m_\omega^2)\Bigr]A_0(m_\omega^2)\nonumber\\
	&&-(2-D)q^2(q^2-4m_\rho^2)\Bigl[8(D-1)m_\rho^6-4(2D-3)(M^2+q^2-m_\omega^2)m_\rho^4\nonumber\\
	&&-q^2\bigl((10-7D)M^2-2(D-1)q^2+(7D-10)m_\omega^2\bigr)m_\rho^2\nonumber\\
	&& -(D-2)q^4(M^2-m_\omega^2)\Bigr]A_0(M^2)-q^2m_\rho^2\Bigl[-16(D-1)\bigl(2(D-1)M^2 \nonumber\\
	&&+(D-4)q^2-2(D-1)m_\omega^2\bigr)m_\rho^6\nonumber\\
	&&+4\Bigl(4(D^2-3D+3)M^4+4(3D-5)q^2M^2 \nonumber\\
	&&+(8D^2-37D+33)q^4+4(D^2-3D+3)m_\omega^4-4\bigl(2(D^2-3D+3)M^2 \nonumber\\
	&&+(4D^2-13D+11)q^2\bigr)m_\omega^2\Bigr)m_\rho^4\nonumber\\
	&&-2q^2\Bigl(2(D^2-5D+8)M^4 +(-7D^2+40D-49)q^2M^2\nonumber\\
	&&+(5D^2-28D+31)q^4+2(D^2-5D+8)m_\omega^4\nonumber\\
	&&-\bigl(4(D^2-5D+8)M^2+(9D^2-40D+47)q^2\bigr)m_\omega^2\Bigr)m_\rho^2 \nonumber\\
	&&+q^4\Bigl((D^2-2D+4)M^4-2(D^2-8D+10)q^2M^2\nonumber\\
	&&+(D^2-6D+8)q^4 +(D^2-2D+4)m_\omega^4-2\bigl((D^2-2D+4)M^2 \nonumber\\
	&&+(D^2-4D+6)q^2\bigr)m_\omega^2\Bigr)\Bigr]B_0(q^2,M^2,m_\omega^2)\nonumber\\
	&&+m_\rho^2\Bigl[4(D-2)M^2q^8-m_\rho^2\Bigl(-4(D-1)Dm_\omega^4\nonumber\\
	&&+4(D-1)\bigl(D(2M^2+q^2)-q^2\bigr)m_\omega^2 \nonumber\\
	&&-4D^2(M^4-M^2q^2)+D(4M^4+q^4)-2(q^4+10M^2q^2)\Bigr)q^4 \nonumber\\
	&&+32(D^2-3D+2)m_\rho^8q^2\nonumber\\
	&&+2m_\rho^4\Bigl(-8(D-1)^2M^4+2(11D^2-30D+19)q^2M^2 \nonumber\\
	&&+(2D^2+D-5)q^4-8(D-1)^2m_\omega^4+2(D-1)\bigl(8(D-1)M^2 \nonumber\\
	&&+(5D-13)q^2\bigr)m_\omega^2\Bigr)q^2\nonumber\\
	&&+16(D-1)m_\rho^6\Bigl(2(D-1)M^4+(9-5D)q^2M^2-(D-1)q^4 \nonumber\\
	&&+2(D-1)m_\omega^4+\bigl((7-3D)q^2-4(D-1)M^2\bigr)m_\omega^2\Bigr)\Bigr]B_0(m_\rho^2,M^2,M^2) \nonumber\\
	&&+\Bigl[-32(D^2-3D+2)q^2m_\rho^{10}\nonumber\\
	&&-16(D-1)\Bigl(2(D-1)M^4-3(D-1)q^2M^2+(5-2D)q^4\nonumber\\
	&& +2(D-1)m_\omega^4+\bigl((13-5D)q^2-4(D-1)M^2\bigr)m_\omega^2\Bigr)m_\rho^8 \nonumber\\
	&&-2q^2\Bigl(-16(D-1)M^4+22(D-1)^2q^2M^2+(4D^2-19D+17)q^4\nonumber\\
	&&-16(D-1)m_\omega^4 +2(D-1)\bigl(16M^2+(21D-53)q^2\bigr)m_\omega^2\Bigr)m_\rho^6 \nonumber\\
	&&+q^4\Bigl(4(7D^2-25D+18)M^4+2(9D^2-24D+19)q^2M^2 \nonumber\\
	&&+(D-2)^2q^4+4(7D^2-25D+18)m_\omega^4\nonumber\\
	&&+\bigl(2(11D^2-44D+37)q^2-8(7D^2-25D+18)M^2\bigr)m_\omega^2\Bigr)m_\rho^4 \nonumber\\
	&&-2q^6\Bigl((5D^2-19D+16)M^4+(D-2)^2q^2M^2 +(5D^2-19D+16)m_\omega^4\nonumber\\
	&&+\bigl((D-2)^2q^2-2(5D^2-19D+16)M^2\bigr)m_\omega^2\Bigr)m_\rho^2\nonumber\\
	&&+(D-2)^2q^8(M^2-m_\omega^2)^2\Bigr]B_0(m_\rho^2,M^2,m_\omega^2) \nonumber\\
	&&-2(D-1)m_\rho^2\Bigl[4M^2(M^2-m_\omega^2)q^8\nonumber\\
	&&-m_\rho^2\Bigl(2Dm_\omega^6+\bigl(2q^2-3D(2M^2+q^2)\bigr)m_\omega^4 \nonumber\\
	&&+\bigl(D(6M^4+2q^2M^2+q^4)-2(q^4+10M^2q^2)\bigr)m_\omega^2\nonumber\\
	&&+M^2\bigl(6(3M^2-2q^2)q^2 +D(-2M^4+q^2M^2+q^4)\bigr)\Bigr)q^4\nonumber\\
	&&+8m_\rho^8\bigl(2(D-1)M^2+(D-4)q^2 -2(D-1)m_\omega^2\bigr)q^2\nonumber\\
	&&+m_\rho^4\Bigl(-8(D-1)M^6+2(9D+7)q^2M^4+(13D-68)q^4M^2 \nonumber\\
	&&+(D-2)q^6+8(D-1)m_\omega^6+\bigl(2(15-7D)q^2-24(D-1)M^2\bigr)m_\omega^4 \nonumber\\
	&&+\bigl(24(D-1)M^4-4(D+11)q^2M^2+(3D-20)q^4\bigr)m_\omega^2\Bigr)q^2 \nonumber\\
	&&+2m_\rho^6\Bigl(8(D-1)M^6-16Dq^2M^4+2(24-7D)q^4M^2\nonumber\\
	&&+(9-2D)q^6 -8(D-1)m_\omega^6\nonumber\\
	&&+\bigl(-24(D-1)M^4+8\bigl(3(D-1)M^2+2(D-2)q^2 \nonumber\\
	&&+32q^2M^2-2(D-8)q^4\bigr)m_\omega^2\bigr)m_\omega^4\Bigr)\Bigr]C_0(m_\rho^2,m_\rho^2,q^2,M^2,M^2,m_\omega^2)
	\Bigr\},\\ [10pt]
	f_3^{(21+22)}(q^2) &=&0.
	\en

	\section{Regular parts}
	\label{app:regular}
	
	In this appendix, we list the regular parts of the integrals $I_\alpha$, $\alpha=1,\ldots,13$\,,
	calculated up to and including $O(q^6)$
	\eq
	I_1^R
	&=&
	-i\pi\left(1+2M^2+\frac{q^2}{6}+6M^4+\frac{q^4}{30}\right)
	\nonumber\\[2mm]
	&+&
	\hat L\left(2+4M^2+\frac{q^2}{3}+12M^4+\frac{q^4}{15}\right)
	\nonumber\\[2mm]
	&-&\left(4M^2-\frac{q^2}{6}+16M^4-\frac{M^2q^2}{3}-\frac{q^4}{20}\right)\, ,
	\\[4mm]
	I_2^R
	&=&i\pi\,\left(-1+\left(M^2-\frac{q^2}{6}+\Delta\right)
	+\left(3M^4-q^2M^2-\frac{q^4}{30}-\Delta^2\right)\right)
	\nonumber\\[2mm]
	&-&\hat L\left(\left(3M^2+\Delta\right)+\left(7M^4+q^2M^2-2M^2\Delta-\Delta^2\right)\right)
	\nonumber\\[2mm]
	&+&\left(\left(M^2+\frac{q^2}{2}\right)
	+\left(-\frac{\Delta^2}{2}-\frac{\Delta M^2}{2}+\frac{29 M^4}{6}
	+M^2q^2-q^2\Delta+\frac{q^4}{4}\right)\right)\, ,
	\\[4mm]
	I_3^R
	&=&\hat L\left(1+\frac{q^2}{6}+\frac{q^4}{30}\right)
	\nonumber\\[2mm]
	&+&\left(1+\left(-\frac{M^2}{3}+\frac{2q^2}{9}+\frac{\Delta}{2}\right)
	+
	\left(-\frac{\Delta^2}{6}+\frac{M^2\Delta}{6}+\frac{q^2\Delta}{12}-\frac{M^4}{20}
	-\frac{M^2q^2}{15}+\frac{23q^4}{450}\right)\right)\, ,
	\nonumber\\
	\\
	I_4^R&=&
	-\frac{\sqrt{3}\pi}{9}
	-\left(-\frac{1}{9}+\frac{5\sqrt{3}\pi}{162}\right)q^2
	-\left(-\frac{7}{180}+\frac{7\sqrt{3}\pi}{810}\right)q^4\, ,
	\\
	I_5^R&=&-2\hat L+2-\frac{\sqrt{3}\pi}{3}\, ,
	\\
	I_6^R&=&-2\hat L+\frac{q^2}{6}+\frac{q^4}{60}\, ,
	\\
	I_7^R &=&0\, ,
	\\
	I_8^R&=&\hat L\left(-2+M^2-\Delta\right)
	+\left(2+M^2-\frac{M^4}{6}+\frac{\Delta M^2}{2}-\frac{\Delta^2}{2}\right)\, ,
	\\
	I_9^R
	&=& -2\hat L\left(1+M^2+\left(M^4+q^2M^2-\Delta M^2\right) \right)\nonumber\\[2mm]
	&+&\left(1+\left(M^2-\Delta+\frac{q^2}{2}\right)+\left(M^4-2\Delta M^2+\frac{5M^2q^2}{2}+\frac{q^4}{6}- \frac{\Delta q^2}{2}+\frac{\Delta^2}{2}\right)
	\right)\, ,
	\\
	I_{10}^R
	&=&-2\hat L(1-2M^2-2M^4)+i\pi(1-2M^2-2M^4)+2(1-2M^4)\, ,
	\\
	I_{11}^R&=&0\, ,
	\\
	I_{12}^R&=&-2\hat L+1\, ,
	\\
	I_{13}^R&=&-2\hat L(1+\Delta)+\left(1-\frac{\Delta^2}{2}\right)\, .
	\en
	
	\section{The self-energy of the $\rho$-meson}
	\label{app:SE}
	
	In this appendix, following
	Ref.~\cite{Djukanovic:2009zn}, we consider the use of  
	the complex-mass renormalization scheme (CMS) for
	calculating the counterterms $\delta m^2_\rho$ and  $\delta c_x$, and
	show that this prescription is equivalent to the IR. For completeness,
	we also present the result for the wave-function renormalization constant. 
	
	Using Eq.~(\ref{eq:SE}), it is seen that only $\Pi_1(p^2)$ contributes to the complex pole position of the two-point function 
	\eq
	s_R-m_{\rho,0}^2-\Pi_1(s_R)=0\,.
	\en
	Here, $m_{\rho,0}$ is the bare mass and $\Pi_1(s_R)$ is UV-divergent. We split the bare mass into the finite renormalized mass and the counterterm $m_{\rho,0}^2=m^2_\rho+\delta m_\rho^2$.
	In the complex-mass scheme, we {\em choose} the renormalized mass to be the complex pole position $s_R$ of the dressed propagator in the chiral limit. Furthermore, we fix the counterterm $\delta c_x$ by demanding that the renormalized parameter $c_x$ is the coefficient of the $O(M^2)$ term in the pion-mass expansion of $s_R$. Writing schematically the expansion in the pion mass of $\Pi_1(p^2)$ as
	\eq
	\Pi_1(p^2)=\Pi^{(0)}_1(p^2)+M^2\Pi^{(1)}_1(p^2)+\ldots,
	\en
	the above prescription amounts to setting
	\eq
	\delta m_\rho^2 &=&-\Pi^{(0)}_1(s_R),\\[2mm]
	\delta c_x &=&-\Pi^{(1)}_1(s_R).
	\en
	\begin{figure}[H]
		\centering
				\includegraphics[width=15cm]{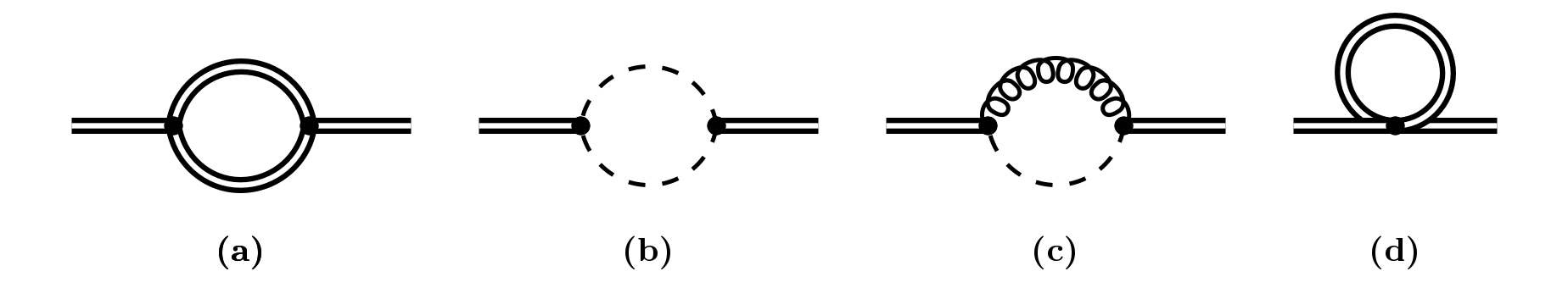}
		\caption{One-loop diagrams contributing to the $\rho$-meson self-energy. Wavy, squiggly, dashed and  double lines correspond to the photon, $\omega$,
			$\pi$ and $\rho$-mesons, respectively.}
		\label{fig:SE}
	\end{figure}
	
	The one-loop diagrams contributing to the $\rho$-meson self-energy up to order $O(q^3)$ are given in Fig.~\ref{fig:SE}. The results for the parts contributing to $\Pi_1(p^2)$ are as follows:
	\eq
	D_a&=&\frac{1 }{128 (D-1) D \pi^2F^2 m_\rho^2}\Bigl\{2 \left(D p^4+4 (D-2) D p^2 m_\rho^2+4 (D-1)m_\rho^4\right)A_0(m_\rho^2)\nonumber\\[2mm]
	&&-D (p^2-4 m_{\rho
	}^2) \left(p^4+4 (2 D-3) p^2 m_\rho^2+4 (D-1) m_\rho^4\right)B_0(p^2,m_\rho^2,m_\rho^2)\Bigr\},\\[2mm]
	D_b&=&\frac{m_\rho^2 \left(2 A_0(M^2)+(4 M^2-p^2)
		B_0(p^2,M^2,M^2)\right)}{32\pi ^2 F^2 (D-1) },\\[2mm]
	D_c&=&-\frac{(D-2)(D-3) g_{\omega \rho \pi }^2 }{64 (D-1) F^2 \pi ^2}\Bigl\{(M^2-p^2-m_{\omega }^2)A_0(m_{\omega }^2)- (M^2+p^2-m_{\omega
	}^2)A_0(M^2)\nonumber\\[2mm] 
	&&+ \left(p^4+(M^2-m_{\omega
	}^2)^2-2 p^2 (M^2+m_{\omega }^2)\right)B_0(p^2,M^2,m_{\omega }^2)\Bigr\},\\[2mm] 
	D_d&=&-\frac{(D-1)^2 m_\rho^2 A_0(m_\rho^2)}{16 D \pi ^2F^2}.
	\en
 	Setting the renormalization scale $\mu=m_\rho$, we obtain the following results for the counterterms:  
	\eq
	\delta m_\rho^2&=&\frac{m_\rho^4}{192 \pi ^2 F^2} (8 g_{\omega \rho \pi }^2+65)\hat{L}+\frac{g_{\omega \rho \pi }^2 m_\rho^4}{36 \pi ^2 F^2}+\frac{(-586+99 \sqrt{3}+12 i \pi ) m_\rho^4}{1152 \pi ^2 F^2},\\[2mm]
	\delta c_x&=& \frac{m_\rho^2}{16F^2\pi^2}(g_{\omega \rho \pi }^2+2)\hat{L}-\frac{ (2+i\pi ) m_\rho^2}{16 F^2 \pi ^2}.
	\en 
	
	Let us see what happens in IR. Subtracting the relevant regular parts given in appendix \ref{app:regular} from the integrals in $D_a$, $D_b$, $D_c$ and $D_d$, one finds:
	\eq
	\Pi^{(0)}_1(s_R)=\Pi^{(1)}_1(s_R)=0\,.
	\en 
	This shows that, in fact, for the self-energy, at given order, the IR and CMS are equivalent procedures. 
	For completeness, we also present here the result for the $Z$-factor in IR (setting
	$\mu=m_\rho=1$):
	\eq
	\delta Z_\rho &=& \frac{\hat{L}}{96\pi^2F^2} \Bigl(g_{\omega \rho \pi }^2 (-M^6+3 (M^4-1) m_{\omega }^2-3 M^2 m_{\omega }^4-3 M^2+m_{\omega }^6+2)\nonumber\\ [2mm]
	&&+4(2 M^2+3) M^4\Bigr)\nonumber\\[2mm]
	&&+\frac{1}{576 \pi ^2 F^2}g_{\omega \rho \pi}^2 \Bigl(
        M^8-12 M^6-11
	M^4+(4-9 M^2) m_{\omega}^6\nonumber\\[2mm]
	&&+6 M^2 \bigl( \log(M^2)-1\bigr) (M^2-m_{\omega }^2-2)+6 m_{\omega}^2 (-M^2+m_{\omega}^2-2) \bigl( \log(m_{\omega}^2)-1\bigr)-30 M^2
        \nonumber\\[2mm]
        &&+6 (M^4+(1-2 M^2) m_{\omega }^2+M^2+m_{\omega
	}^4-2)\Bigl\{
        (M^2-m_{\omega }^2-1) \log \left(\frac{m_{\omega }}{M}\right)
	+2- \log (M^2)
        \nonumber\\[2mm]
        &&-2 M m_{\omega } \sqrt{1-\frac{(\Delta +M^2)^2}{4 M^2 m_{\omega }^2}} \mbox{arccos} \left(\frac{\Delta +M^2}{2 M m_{\omega }}\right)\Bigr\}
        \nonumber\\[2mm]
	&&+2 (5M^4-9 M^2-6) m_{\omega }^4+(-5 M^6+28 M^4+9 M^2-30) m_{\omega }^2+3 m_{\omega }^8+29\Bigr)\nonumber\\[2mm]
	&&+\frac{1}{288 \pi ^2 F^2}\bigl(3 \log (M^2)-3 (2 M^2 +1 )\sigma \log \left(\frac{1-\sigma }{\sigma +1}\right)\nonumber\\[2mm]
	&&+6M^2-18M^4-32M^6+3i\pi(1-\sigma-2\sigma M^2-6M^4)\bigr)\,.
	\en

	\section{Behavior near $q^2=0$}
	\label{app:q2}

	\begin{figure}[t]
		\begin{center}
			\includegraphics[width=5.5cm]{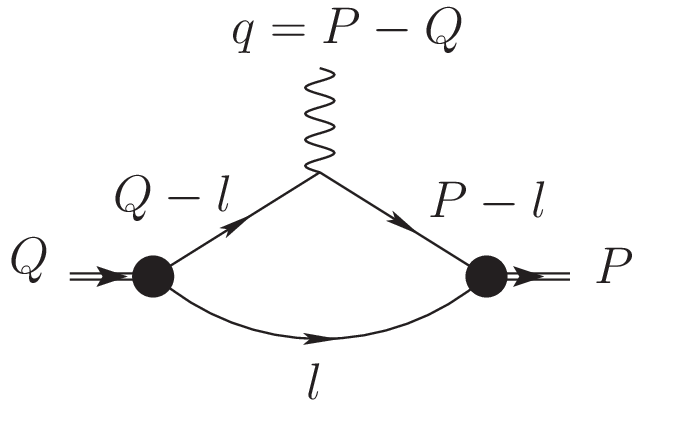}
			\caption{Triangle diagram contributing to the $\rho$-meson form factors. Calculations are
				done in the Breit frame, where $\bm{P}=-\bm{Q}=\bm{q}/2$. The filled circles
				denote the $\rho\pi\pi$ vertices.}
			\label{fig:triangle_rho}
		\end{center}
	\end{figure}

	As mentioned above, the resonance form factors, calculated in NREFT, demonstrate rapid
	variations in the vicinity of $q^2=0$. It would be very interesting to understand, in simple terms, where such a behavior comes from. Moreover, it would be intriguing to know,
	whether this behavior is universal, or characterizes the case of the $\rho$-meson only.
	To answer these questions, consider, for example, the form factor $f_1(q^2)$. Using Eqs.~(3.51)-(3.56) of Ref.~\cite{Meissner:2026zos}, one may write
	down a symbolic expression for the contribution of the triangle graph, see Fig.~\ref{fig:triangle_rho}:
	\eq\label{eq:f1}
	f_1^{\sf triangle}(q^2)=Z\langle F_1\rangle=\int\frac{d^d\bm{l}}{(2\pi)^d}\,
	(\bm{l}^2-(\bm{n}{\bm l})^2)f(\bm{l}_P^2)f(\bm{l}_Q^2)A(\bm{l},q)D(\bm{l},q)\, .
	\en
	The calculation is performed in the Breit frame, where $\bm{P}=-\bm{Q}=\bm{q}/2$
	and $\bm{q}^2=-q^2$. The unit vector $\bm{n}=\bm{q}/|\bm{q}|$ and $Z$ is
	the wave function renormalization constant of the $\rho$-meson in the NREFT framework.
	Furthermore, $D(\bm{l},q)$ denotes the product of two energy denominators. The explicit
	expression, given in Ref.~\cite{Meissner:2026zos}, will not be needed here. Near $q^2=0$,
	it can be expanded in powers of $\bm{q}$:
	\eq 
	D(\bm{l},q)=\frac{1}{(\bm{l}^2-q_R^2)^2}-\frac{q^2}{2s_R}\left((\bm{n}{\bm l})^2-
	\frac{s_R}{4}\right)\frac{1}{(\bm{l}^2-q_R^2)^3}
	-\frac{q^2}{4}\,(\bm{n}{\bm l})^2\frac{1}{(\bm{l}^2-q_R^2)^4}
	+O(q^4)\, ,\quad\quad
	\en
	where $q_R^2=\dfrac{s_R}{2}-M^2$ and $s_R$ denotes the position of the $\rho$-meson
	pole on the complex $s$-plane. Next, $A(\bm{l},q)$ is a low-energy polynomial, with a typical hard scale $M$. Its explicit form does not play any role in the discussion.
	
	A crucial difference between NREFT and the calculations carried out in this paper
	consists in the presence of the $\rho\pi\pi$ vertices $f(\bm{l}_{P,Q}^2)$ in Eq.~(\ref{eq:f1}). The function $f(q^2)$ is determined through the P-wave phase shift via the matching condition~\cite{Meissner:2026zos}
	\eq
	f^2(q^2)=24\pi\sqrt{s}\,\frac{\tan\delta(q)}{q^3}\, ,\quad\quad
	s=4(M^2+q^2)\, .
	\en
	The square of relative three-momenta in the vertices at small $q^2$ are given by
	\eq
	\bm{l}_P^2&=&\bm{l}^2-\frac{|\bm{q}|}{2}(\bm{n}\bm{l})
	-\frac{q^2}{16}\left(1-\frac{(\bm{n}\bm{l})^2}{M^2}\right)+\cdots\, .
	\nonumber\\[2mm]
	\bm{l}_Q^2&=&\bm{l}^2+\frac{|\bm{q}|}{2}(\bm{n}\bm{l})
	-\frac{q^2}{16}\left(1-\frac{(\bm{n}\bm{l})^2}{M^2}\right)+\cdots\, .
	\en
	so that 
	\eq
	f(\bm{l}_P^2)f(\bm{l}_Q^2)
	&=&f^2(\bm{l}^2)-\frac{q^2}{8}\,f(\bm{l}^2)f'(\bm{l}^2)\left(1-\frac{(\bm{n}\bm{l})^2}{M^2}\right)
	\nonumber\\[2mm]
	&+&\frac{q^2}{4}\,((f'(\bm{l}^2))^2-f(\bm{l}^2)f''(\bm{l}^2))(\bm{n}\bm{l})^2
	+O(q^4)\, .
	\en
	In order to integrate in dimensional regularization, one has to average over directions,
	replacing, e.g., $l_il_j\to\dfrac{1}{d}\,\delta_{ij}\bm{l}^2$, and so on. Next, one should expand the numerator at $\bm{l}^2=q_R^2$ in Taylor series. This, in particular, implies
	that the function $f(\bm{l}^2)$ and its derivatives, appearing in the numerator, should be also expanded 
	\eq
	f^{(n)}(\bm{l}^2)=f^{(n)}(q_R^2)+(\bm{l}^2-q_R^2)f^{(n+1)}(q_R^2)+\cdots\, ,
	\quad\quad n=0,1,\cdots\, .
	\en
	The expansion of the numerator is truncated, when all denominators,
	containing powers of $(\bm{l}^2-q_R^2)$, are canceled.
	The remaining integrals can be easily done analytically in dimensional regularization.
	The large numerical factors, which were mentioned in the text, emerge from the derivatives $f^{(n)}(q_R^2)$. In the vicinity of a resonance, the phase shift can be approximated as
	follows
	\eq
	\tan\delta(q)\simeq \frac{C}{s-\mbox{Re}\,s_R}\, ,\quad\quad s\to \mbox{Re}\,s_R\, .
	\en
	Consequently,
	\eq
	f(q^2)\propto (s-\mbox{Re}\,s_R)^{-1/2}\, ,\quad\quad
	f^{(n)}(q^2)\propto (s-\mbox{Re}\,s_R)^{-1/2-n}\, ,
	\en
	and
	\eq
	f^{(n)}(q_R^2)\propto (\mbox{Im}\,s_R)^{-1/2-n}\, .
	\en
	It is seen that the role of the small scale is played by $\mbox{Im}\,s_R$, i.e., by the decay width. Hence, for example, the resonance charge radius diverges in the limit of the vanishing width.

To summarize, we have demonstrated that the emergence of a small scale in the
resonance form factor is a non-perturbative effect, since the pole in the function
$f(q^2)$ cannot be obtained perturbatively at any order. In addition, we note that in the
derivation we have not used any specific properties of the $\rho$-meson (for example,
the fact that it has spin one). Hence, the main result -- the appearance of a small scale
in the form factors -- is general and applies to any low-lying narrow resonance. Kinematic
effects related to the spin may further enhance this phenomenon, as, for example, in
the case of the quadrupole moment of a spin-one resonance. In our opinion, it would
be extremely interesting to verify this effect, either in the experiment or on the lattice.
Note also that the above argumentation uses the dimensional
regularization. However, it is expected that physical results should not depend on the use
of a particular regularization.

	\bibliographystyle{unsrt}
	\bibliography{ref}

\end{document}